\documentclass[twocolumn,english,showpacs,aps,pre,reprint,floatfix]{revtex4-2}

\usepackage{geometry}
\usepackage{ragged2e}
\usepackage{xcolor}
\usepackage{babel}
\usepackage{units}
\usepackage{amsmath}
\usepackage{amssymb}
\usepackage{stmaryrd}
\usepackage{graphicx}
\usepackage{wasysym}
\usepackage{bbold}
\usepackage{mathtools}
\usepackage{blkarray, bigstrut}

\usepackage{wrapfig}
\newcommand{\DQC}{Duke Quantum Center, Departments of Electrical and Computer Engineering and Physics, \\Duke University, Durham, NC 27708}

\begin{document}
\title{Heating and Escape of Confined Atoms Subject to Colored Noise}
\date{\today}

\author{Joseph M. Ryan}
\email{joseph.ryan@duke.edu}
\author{Katherine Jonas}
\author{Christopher Monroe}
\email{c.monroe@duke.edu}
\affiliation{\DQC}

\begin{abstract}
The motion of a particle in a finite-depth confining potential, such as that of an electromagnetically trapped atom or ion, is approximately harmonic at low energy and softens as the escape energy is approached.  When the particle is driven by a stochastic force, the heating and escape times depend sensitively on the noise power spectrum.  We study escape times of particles in softening potentials driven by power-law colored noise.  We identify two nested conditions for weak noise: the particle must execute many oscillations before escape, and noise forces must be small relative to forces exerted by the trap so as not to tilt the potential appreciably.  In this regime, we derive expressions for the mean escape time using stochastic averaging techniques and compare them with Monte Carlo simulations of the full equations of motion.  We also examine the crossover to regimes where these conditions are not met and large quasi-static forces give rise to additional escape mechanisms.  Applying these results to an experimental measurement of lifetimes in a microfabricated ion trap, we find that spectral color alone cannot account for the rapid escape rates observed, whereas quasi-static barrier tilting by a stray field of plausible magnitude could.  These results are relevant to the heating and escape of trapped atoms due to noisy force fields, as well as to other nonlinear oscillators subject to colored noise.
\end{abstract}

\maketitle
\section{Introduction}\label{section:introduction}
Confinement dynamics in anharmonic potentials subject to stochastic forces touch on many areas of physics~\cite{Hanggi1990, Raychaudhuri2002} and are increasingly relevant to quantum technologies, from trapped ions~\cite{Home_2011} and atoms~\cite{Grimm_2000, Gibbons_2011} to superconducting systems~\cite{Paladino2014}.  Colored noise forcing in particular can give rise to unusual behavior, such as anomalous diffusion. Because it has frequency-dependent spectral power, its effect depends on the frequencies sampled by the nonlinear dynamics~\cite{Mallick_2002,Mallick_2004, Dykman_1997, Hanggi_colored_1988}. 

% Here, the ``softening'' of the oscillation at higher energies hastens the heating and escape times.

%subject to colored noise,The simple case of a single particle in a smooth potential with finite depth exhibits negative anharmonicity at high energies, causing the oscillation frequency to decrease as energy increases.   This property, often dubbed the `softening' spring, becomes particularly important when the particle is subject to noise, as the resulting dynamics of such nonlinear oscillators can depend sensitively on the power spectrum of the noise.

Fluctuating electric fields originating from ion trap surfaces can heat trapped ions until they eventually escape.  An experiment with a very small ion trap~\cite{Stick2006} displayed escape times substantially shorter than what would be expected from a white noise source (having a flat spectrum independent of frequency), given the measured low-energy heating rate and trap depth.  One possible explanation is colored noise with increasing spectral density at lower frequencies, which hastens escape in a ``softening'' trap.

Both the mean escape rate $\dot E_{\mathrm{esc}}$ (expressed in terms of energy per unit time) and the heating rate near the bottom of the confining potential $\dot E_{\mathrm{cold}}$ (where the trap is approximately harmonic) are independently measurable quantities that are sensitive to the noise spectrum, albeit to different frequency components. Because they are both proportional to the intensity of the noise source, we introduce the dimensionless normalized escape time  
\begin{equation}\label{eq:ratio_definition}
    \tau_{\mathrm{esc}}\equiv\frac{\dot E_{\mathrm{cold}}}{\dot E_{\mathrm{esc}}}
= \frac{T_1 \dot E_{\mathrm{cold}}}{\Delta}.
\end{equation}
Here, $\dot E_{\mathrm{esc}}=\Delta / T_1$, where $\Delta$ is the depth of the confining potential and $T_1$ is the mean escape time for particles starting at rest at the bottom of the well.  The escape rate $\dot{E}_{\mathrm{esc}}$ is rarely reported, mainly because of the very long wait times necessary for such a measurement. 

The linear heating rate $\dot E_{\mathrm{cold}}$ is routinely measured, and its relationship to the force noise at the harmonic oscillation frequency $\omega_0$ is
\begin{equation} \label{eq:Ecold}
    \dot E_{\mathrm{cold}}=\frac{S_\xi(\omega_0)}{2m},
\end{equation}
where $S_\xi(\omega)$ is the two-sided power spectral density (PSD) of the stochastic force at frequency $\omega$~\cite{Savard1997,Turchette2000,Brownnutt2015} and $m$ is the mass of the confined atom. For a trapped ion with charge $e$, the heating force noise is $S_\xi(\omega)=e^2 S_\text{E}(\omega)$, where $S_\text{E}(\omega)$ is the electric field noise PSD.
For example, thermal noise on electrodes at characteristic distance $d$ from a trapped ion gives rise to (white) electric field noise $S_\text{E}(\omega)\sim 2k_BTR/d^2$, where $T$ is the electrode temperature and $R$ is the effective (lumped-circuit) resistance between trap electrodes~\cite{Turchette2000}, valid in the near-field limit where $d \ll c/\omega_0$ and $c$ is the speed of light. Analogous cold heating rates for optically confined neutral atoms subject to laser pointing noise can also be considered~\cite{Savard1997, Gehm_1998}.

The normalized escape time $\tau_\text{esc}$ quantifies how noise color and the trap anharmonicity jointly govern escape.  In the white-noise case, $\dot E_{\mathrm{cold}}$ is independent of trap frequency and $\tau_{\text{esc}}=1$. Our objective is to understand how colored noise, particularly noise that grows at lower frequencies, such as $1/f$ noise, modifies $\tau_\text{esc}$ from unity.

\section{System description}\label{section:system_description}
\begin{figure}
    \centering
    \includegraphics[width=1\linewidth]{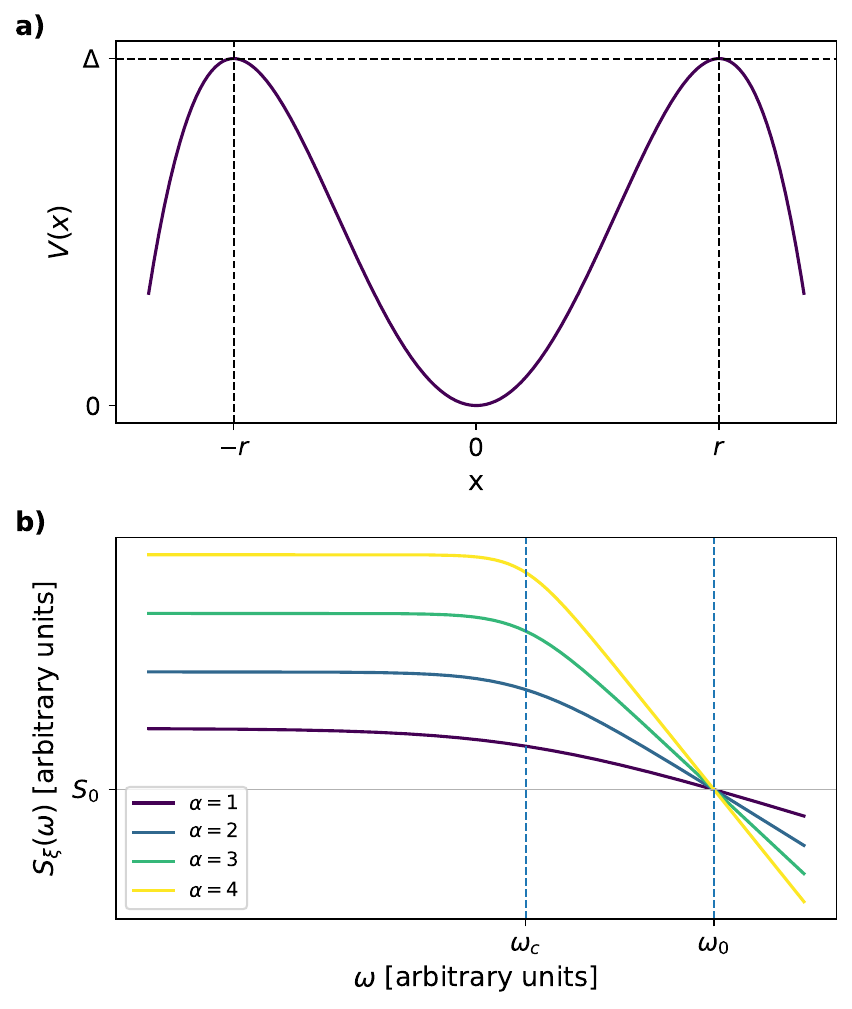}
    \caption{(a) Potential $V(x)$ from~Eq.~\eqref{eq:potential}.  (b)  Power spectral density $S_\xi(\omega)$ in the form of~Eq.~\eqref{eq:power_law_definition_with_cutoff}, with low-frequency cutoff $\omega_c$.  Note that for different values of $\alpha$, we normalize the PSDs such that they all have the same value $S_0 = S_\xi(\omega_0)$.  With this normalization choice, $\dot E_\text{cold}$ is identical for all values of $\alpha$.}
    \label{fig:potpsd}
\end{figure}

We model the potential $V(x)$ of a finite-depth trap in one dimension as a harmonic potential with a negative quartic term, having parameters $r$ and $\Delta$
\begin{equation}\label{eq:potential}
V(x) = 2\Delta \,\frac{x^2}{r^2}\left(1 - \frac{x^2}{2r^2}\right).
\end{equation}
Here $x$ is the distance from the trap center, $r$ is the distance from the trap center to either potential maximum, and $\Delta$ is the trap depth.  This potential, illustrated in~Fig.~\ref{fig:potpsd}(a), has a minimum at $x=0$, maxima at $x=\pm r$, and is anti-trapping for $|x|>r$.

Any potential of the form $V(x) = (1/2)ax^2 - (1/4)bx^4$ can be mapped to~Eq.~\eqref{eq:potential} because there is a one-to-one reparameterization map $(a,b)\leftrightarrow(\Delta,r)$.  We choose to use $\Delta, r$ because they are natural parameters in ion-trap experiments.  Furthermore, by rescaling $x = r\tilde x$ and $t = \tilde t/\omega_0$, it is possible to map the dynamics in any of these potentials to dynamics in the potential $V(\tilde x) = (1/2)\tilde x^2 - (1/4)\tilde x^4$.  Therefore, there is no free ``shape'' parameter.  The rescaling applies to the noise too, so the escape problem depends on the trap only through the noise spectrum expressed in trap units.  

We assume that the ion is driven with a stochastic force $\xi(t)$.  The Langevin equation representing this process is
\begin{equation}\label{eq:duffing}
m\ddot{x} + m\omega_0^2x\Big(1-\frac{x^2}{r^2}\Big) = \xi(t),    
\end{equation}
where $\omega_0^2 \equiv V''(0)/m = 4\Delta/mr^2$ is the small-amplitude (linear) trap frequency.  We do not consider damping, as most atom trap experiments have little damping in the absence of laser cooling.  We define the escape time as the first time the instantaneous energy $E(t) = m\dot x^2/2 + V(x)$ reaches $\Delta$.  

The noise force $\xi(t)$ is often taken to follow a pure power-law PSD $S_\xi(\omega) \propto 1/\omega^{\alpha}$ as reported in many atom trap measurements, with wide variations of the exponent $\alpha$.  Typically, trapped-ion experiments measure this exponent over a frequency range $100\text{ kHz }\lesssim\omega/2\pi\lesssim10\text{ MHz}$ and report values of $1\lesssim\alpha\lesssim 3$~\cite{Turchette2000,Brownnutt2015}. Pure power-law noise is ubiquitous in nature, but also has a divergent total power.  It therefore requires infrared regularization for $\alpha \geq 1$, and ultraviolet regularization for $\alpha \leq 1$.  We include an infrared rolloff $\omega_c$ explicitly, below which the noise becomes effectively white.  The interpretation of low-frequency cutoffs and the origin of $1/f$ noise are themselves well-studied in many systems~\cite{Barkai_2013, Dutta_1981}.  Some studies have probed low-frequency electric field noise in ion traps~\cite{Wong-Campos2016, Bonus2025}.  We also assume that there is an unimportant high-frequency cutoff above the dynamically relevant frequencies, which we do not model.  We consider spectra of the form
\begin{equation}\label{eq:power_law_definition_with_cutoff}
    S_\xi(\omega) = S_0\left(\frac{\omega_c^\alpha + \omega_0^\alpha}{\omega_c^\alpha + |\omega|^\alpha}\right),
\end{equation} 
with $S_0 \equiv S_\xi(\omega_0)$, which are illustrated in Fig.~\ref{fig:potpsd}(b) for different values of $\alpha$. 

For $\alpha = 2$, the PSD is a Lorentzian function, corresponding to an exponentially decaying autocorrelation function $\langle \xi(t)\xi(t+t')\rangle_t  = \frac{1}{2\pi}\int_{-\infty}^\infty S_\xi(\omega)e^{-i\omega t'}d\omega =  \frac{\sigma^2}{2\omega_c}e^{-|t'|/\tau}$ with correlation time $\tau=1/\omega_c$ and $\sigma^2=S_0(\omega_c^2+\omega_0^2)$. This is the Ornstein--Uhlenbeck (O-U) noise process~\cite{Ornstein_1930}.

\section{Harmonic oscillator driven by Ornstein--Uhlenbeck noise}\label{section:sho_ou}
Before treating the escape problem, it is useful to consider the effect of low-frequency forces in a harmonic trap.  We illustrate this point using an O-U noise process with PSD of the form of~Eq.~\eqref{eq:power_law_definition_with_cutoff} with $\alpha = 2$.  We use this noise process as it has a simple stochastic differential equation representation
\begin{equation}\label{eq:OU_SDE}
    d\xi = -\omega_c\xi dt + \sigma dW_t,
\end{equation}
where $dW_t$ is a standard Wiener process and $\sigma^2 = S_0(\omega_c^2 + \omega_0^2)$ to be consistent with~Eq.~\eqref{eq:power_law_definition_with_cutoff} with $\alpha = 2$.  Furthermore, unlike ideal white noise and $1/f$ noise, it has finite power.

A harmonic oscillator driven by  O-U noise obeys
\begin{align}\label{eq:OU_driven_SHO}
    m\ddot{x} + m\omega_0^2x = \xi(t),
\end{align}
where $\xi(t)$ evolves according to~Eq.~\eqref{eq:OU_SDE}.  We assume that the particle begins at rest at the origin, or $x(0) = 0$ and $\dot{x}(0) = 0$, and that the noise is already in statistical equilibrium so that $\xi(0)$ has a normal distribution with mean $0$ and variance $\sigma^2/(2\omega_c)$. We examine the ensemble-averaged energy
\begin{equation}\label{eq:mean_energy_def}
    \langle E(t) \rangle = \frac{m\omega_0^2 \langle x^2(t) \rangle + m\langle \dot{x}^2(t) \rangle}{2}.
\end{equation}

The conventional treatment of heating in a harmonic potential gives the linear heating rate defined in~Eq.~\eqref{eq:Ecold}.  While this holds in the long-time limit ($t\gg 1/\omega_c$), it fails to capture relevant dynamics at short times.  The Fokker--Planck equation provides the general solution for the mean energy (see Appendix~\ref{appendix:ou_sho})
\begin{align}\label{eq:sho_exact}
    &\langle E(t)\rangle = \frac{\sigma^2}{2m(\omega_0^2 + \omega_c^2)}t + \frac{\sigma^2 (\omega_0^2 - \omega_c^2)}{2m\omega_c(\omega_0^2 + \omega_c^2)^2} \nonumber \\
    &\quad +\frac{\sigma^2e^{-\omega_c t}}{2m\omega_c(\omega_0^2+\omega_c^2)^2} \nonumber\\
    &\quad\times \left[ (\omega_c^2-\omega_0^2)\cos (\omega_0t) - 2\omega_0\omega_c \sin (\omega_0t)\right].
\end{align}     
Removing the exponentially decaying oscillations yields the mean asymptotic energy (for $t \gg \tau$)
\begin{align}
     \langle E(t) \rangle  &\simeq  E_\text{offset} + \frac{1}{2m}S_0 t \nonumber\\
    &\simeq E_\text{offset} +  \dot E_\text{cold} t \label{eq:mean_energy_oscillation_averaged},\\
    E_\text{offset} &= \frac{\sigma^2(\omega_0^2 -\omega_c^2)}{2m\omega_c(\omega_0^2 + \omega_c^2)^2}\label{eq:offset_energy}.
\end{align}

\begin{figure}
    \centering
    \includegraphics[width=1\linewidth]{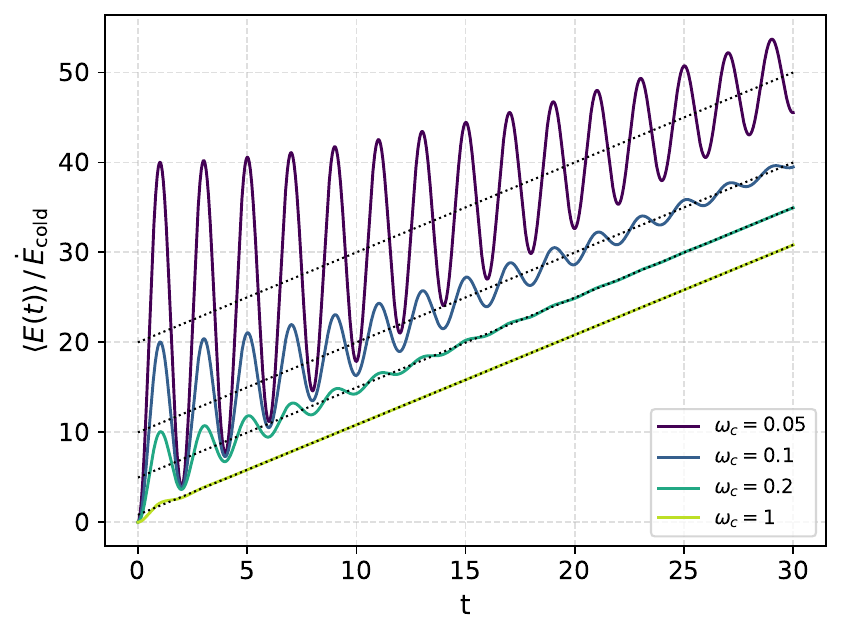}
    \caption{Mean energy of a harmonic oscillator with $\omega_0 = \pi$ driven by O-U noise evaluated using~Eq.~\eqref{eq:sho_exact} and normalized by the cold heating rate $\dot E_\text{cold}$.  We show different values of the low-frequency cutoff $\omega_c$.  All four curves approach unit slope at long times, as required by~Eq.~\eqref{eq:Ecold}.  Dotted lines show the asymptote~Eq.~\eqref{eq:mean_energy_oscillation_averaged}.  Transient oscillations persist for a duration $\sim1/\omega_c$ and smaller $\omega_c$ raises the intercept $E_\text{offset}$.  }
    \label{fig:energy}
\end{figure}
We illustrate $\langle E(t) \rangle$ for matched $S_0$ and different $\omega_c$ in Fig.~\ref{fig:energy}.   The progressively diminishing oscillations persist for a duration on the order of the noise correlation time~$\tau$.  These occur because low-frequency forces displace the equilibrium.  The offset energy survives at long times because smaller $\omega_c$ means more total power at fixed $S_0$.  A particle cooled to the displaced equilibrium instead shows essentially no transient excitation, and for $\omega_c\ll\omega_0$ its offset is half of~Eq.~\eqref{eq:offset_energy}  (Appendix~\ref{appendix:initial_state_prep}).  

% For noises with high low-frequency power (and thus small $\omega_c$), when the particle begins its trajectory at rest and at the center of the well, there is, on average, a large quasi-DC force that displaces the particle away from the center, giving rise to oscillations in the energy as this force simply displaces the equilibrium.  This force persists for a duration similar to the noise correlation time $\tau$, which explains why the oscillations diminish on that timescale.  After this correlation time, memory of the initial offset has disappeared, and the mean energy is free of oscillations.  However, the initial energy offset persists, which is explained by the fact that O-U noises with smaller $\omega_c$ (at fixed PSD at resonance) have more total power.  

This shows that even in a simple linear oscillator, low-frequency components of the power spectrum cannot be ignored.  Extending this analysis to a finite-depth oscillator is not straightforward, as the moments in the nonlinear analogue of~Eq.~\eqref{eq:OU_driven_SHO} couple to progressively higher-order moments.  Approximate non-Gaussian closure techniques truncate the moment hierarchy at a chosen order, but we do not pursue them~\cite{CRANDALL_1980}.

\section{Escape-time distributions}\label{section:fptds}
In the previous section, we considered $\langle E(t)\rangle$, which is the ensemble average of energies across all trajectories at a given time $t$.  It is of limited use once an appreciable fraction of trajectories have escaped, as the energies of these trajectories grow without bound.  Their inclusion in the mean energy therefore obscures the dynamics of the surviving population.  

We characterize the escape rate $\dot{E}_{\mathrm{esc}}$ through the mean time $T_1$ taken for a trajectory starting at rest to first reach energy $\Delta$~\cite{Redner_2001}.  For each realization, the escape time is
\begin{equation}
T=\inf\{t>0:E(t)\geq\Delta\}.
\label{eq:fpt_definition}
\end{equation}
Trajectories terminate at their first crossing and therefore dynamics outside the well are excluded.  The distribution of escape times $T$ is known as the first-passage-time distribution $P^\text{FPT}(t)$.  We characterize $P^\text{FPT}(t)$ by its mean $T_1$ and its median $T_M$.  The dimensionless ratio $\tau_{\rm esc}$ compares the measured mean escape time with the time required to accumulate the trap depth at the cold, harmonic heating rate.

For white noise, a martingale argument gives $T_1$ exactly, for arbitrary anharmonicity.  Transforming~Eq.~\eqref{eq:duffing} to an It\^o system using $\xi(t)dt = \sqrt{S_0}dW$
% \begin{equation}
% dx=v\,dt,\qquad
% dv=-\frac{V'(x)}{m}\,dt+\frac{\sqrt{S_0}}{m}\,dW.
% \label{eq:white_noise_system}
% \end{equation}
and applying It\^o's lemma to the energy gives 
\begin{equation}
dE = \frac{S_0}{2m}\,dt + \sqrt{S_0}\,v\,dW.
\end{equation}
The potential terms cancel and the mean heating rate is $\dot E_\text{cold} = S_0/2m$ at every instant, independent of amplitude and of the shape of $V$.  Furthermore, $E(t)-(S_0/2m)t$ is a local martingale process, since its differential is an It\^o stochastic integral.  Stopping it at the first-passage time $T$ gives~\cite{Roldan_2023}
\begin{equation}
T_1^\text{(white)} = \frac{2m\Delta}{S_0}, \qquad \tau_{\text{esc}} =  1.
\label{eq:martingale_anchor}
\end{equation}
This result provides an important benchmark for the subsequent simulations.  

No analogous martingale result exists for colored noise $\xi$, because it becomes correlated with the velocity $\langle v\xi\rangle \neq 0$.  We treat the colored-noise escape problem in three stages.  We first identify dimensionless parameters separating multiple escape mechanisms, then develop a weak-noise theory, and finally test these predictions using Monte Carlo simulation.

\section{Escape regimes}
As the trap contributes no dimensionless parameter of its own, the noise expressed in trap units controls escape.  Two escape mechanisms compete.  An atom may gain energy incrementally and climb out of the well over many orbits.  However, for $\omega_c\ll \omega_0$, the noise remains correlated over several oscillation periods. These quasi-static forces tilt the potential, substantially lowering one of the two barriers, and open an escape route that is not well captured by gradual heating alone.  Here, we identify the relevant dimensionless parameters governing these regimes and their effects.

\subsection{Dimensionless parameters}\label{section:dimensionless_params}

 The problem is described by five dimensional quantities ($\Delta$, $r$, $S_0$, $\omega_c$, $m$) spanning three physical dimensions, leaving two dimensionless parameter groups (in addition to the spectral exponent $\alpha$). One is $\omega_c/\omega_0$, and the other is
\begin{equation}\label{eq:h_definition}
h \equiv \frac{\dot E_{\rm cold}}{2\Delta\omega_0}
 = \frac{S_0}{m^2r^2\omega_0^3},
\end{equation}
so that $4\pi h$ is the fraction of the trap depth acquired per oscillation period at the cold heating rate.  Every other dimensionless quantity appearing below is a function of these.  In the white-noise limit, $\omega_0T_1^{(\rm white)}=1/(2h)$.  For a 100 meV trap with $\omega_0 = 2\pi\times 1$~MHz and heating rate of $\dot{\bar n} = \dot E_\text{cold}/\hbar\omega_0 = 100$~quanta/s, $h\sim 3\times10^{-13}$, which corresponds to a white-noise escape time of about three days.

\subsection{Barrier tilting}

When $\omega_c\ll\omega_0$ the force is approximately constant over many oscillation periods.  A static force $F$ tilts the potential to $V(x)-Fx$ which, to first order, lowers one of the two barriers to
\begin{equation}\label{eq:lowering}
\Delta_{\rm eff}\simeq\Delta-|F|r .
\end{equation}
If escape proceeds by gradual heating, as opposed to immediate ejection, the escape time is proportional to the barrier height.  We evaluate the fractional barrier reduction $\chi$ produced by a typical force $\xi_{\rm qs}$
\begin{equation}\label{eq:chi_definition}
\chi \equiv \frac{\xi_{\rm qs}\,r}{\Delta}, \: \text{where} \;\;\;
\xi_{\rm qs}^2\equiv\frac{1}{\pi}\int_0^{\omega_0} S_\xi(\omega)d\omega.
\end{equation}
In terms of the two dimensionless parameters, for spectra of the form~Eq.~\eqref{eq:power_law_definition_with_cutoff} we obtain
\begin{equation}\label{eq:chi_exact}
\begin{split}
    \chi^2 &= \frac{16h}{\pi}\left[ 1+ \left(\frac{\omega_c}{\omega_0}\right)^\alpha\right]\int_0^1\frac{dy}{(\omega_c/\omega_0)^\alpha + y^\alpha} \\
    \chi^2  &\simeq \frac{16h}{\alpha\sin(\pi/\alpha)}\left[\frac{\omega_c}{\omega_0}
                    +\Big(\frac{\omega_c}{\omega_0}\Big)^{(1-\alpha)}\right] \\
    &\quad -\frac{16h}{\pi(\alpha-1)} + \mathcal{O}\big(h(\omega_c/\omega_0)^\alpha\big).
\end{split}
\end{equation}
The approximation holds where $\omega_c \ll \omega_0$ and $\alpha > 1$.  At $\alpha = 1$, the expansion is replaced by $\chi^2 \simeq (16h/\pi)\ln(\omega_0/\omega_c)$.  For $\alpha<1$ the integral in~Eq.~\eqref{eq:chi_definition} is dominated by frequencies of order $\omega_0$ rather than by the rolloff, and $\chi^2$ instead saturates at $16h/[\pi(1-\alpha)]$ as $\omega_c\to 0 $.  

Three features of~Eq.~\eqref{eq:chi_exact} are worth noting.  First, $\chi\propto\sqrt{h}$ at fixed $\omega_c/\omega_0$, so barrier tilting is suppressed more slowly than gradual heating itself as the trap is made deeper.  Second, for $\alpha\geq 1$, $\chi$ diverges as $\omega_c\rightarrow0$ at fixed $h$.  Holding the resonant noise power fixed does not hold the total force variance fixed, and the small-$\omega_c$ limit therefore favors direct escape.  For small $\omega_c$, the tilt parameter of this nonlinear oscillator is related to the first-order offset energy of the linear oscillator~Eq.~\eqref{eq:offset_energy}: $E_\text{offset}/\Delta \simeq \chi^2/4$.  Third, the asymptotic expansion in~Eq.~\eqref{eq:chi_exact} assumes $\omega_c\ll\omega_0$ and cannot be continued to large $\omega_c$.  The integral expression in that equation remains valid and gives the finite white-noise limit $\chi^2\to16h/\pi$.  There the noise is effectively flat below $\omega_0$ and the quasi-static picture no longer applies, so $\chi$ should be read as a barrier-tilt parameter only for $\omega_c\ll\omega_0$.  For a static force $F$, the local minimum disappears when $|F|r/\Delta$ exceeds $\chi_\text{crit}= 8/(3\sqrt{3})\approx 1.54$.  

Averaging $\Delta_{\rm eff}$ over the stationary Gaussian distribution of $F$, for which $\langle|F|\rangle=\sqrt{2/\pi}\,\xi_{\rm qs}$, gives
\begin{equation}\label{eq:tilt_correction}
\tau_{\rm esc}\simeq\tau_{\rm esc}^{(0)}
\left(1-\sqrt{2/\pi}\;\chi\right),
\end{equation}
where $\tau_{\rm esc}^{(0)}$ is the value that gradual heating alone would produce.  This estimate~Eq.~\eqref{eq:tilt_correction} is heuristic and makes a number of uncontrolled assumptions, such as treating the force as static over the entire escape and neglecting changes in orbit frequency under the tilt.  We test it against simulations (described below) in Fig.~\ref{fig:ratio}(b).  At a representative point, $h=6.45\times10^{-5}$ and $\omega_c / \omega_0=0.05 / \pi$, the particle completes $\simeq 10^3$ oscillations before escaping.  This would suggest that escape is entirely gradual, and yet $\chi=0.18$ and~Eq.~\eqref{eq:tilt_correction} predicts a $14\%$ reduction of the escape time.  In this example, the static offset is, on average, resampled a large number of times before escape.  

This raises a natural concern that tilting might require reformulating the stopping criterion itself.  However, under a weak static force $F$, the saddle moves to $x_s \simeq r-Fr^2/(8\Delta)$, where the untilted potential energy is $\Delta[1-(Fr/\Delta)^2/16]$.  The energy threshold therefore shifts only at second order in $F$, while the barrier is lowered at
first order~Eq.~\eqref{eq:lowering}.  Throughout we use $E\geq\Delta$ as the stopping criterion.  We compare the resulting escape times with those obtained using position-crossing criteria in~Appendix~\ref{appendix:numerical_simulation_detail}.

\subsection{The weak-noise limit}

The parameters $h$ and $\chi$ measure different physical requirements, but they are not independent. Equation~\eqref{eq:chi_exact} gives $\chi^2 \geq 16h/\pi$ for all $\alpha$ and $\omega_c$, so $\chi \ll 1$ implies $h\ll 1$, while the converse fails.  We therefore define the weak-noise regime by (for $\omega_c\ll \omega_0$)
\begin{equation}\label{eq:weak_noise_criterion}
h \ll 1
\qquad\text{and}\qquad
\chi \ll 1 .
\end{equation}
The first requires that the particle complete many oscillations before escaping, so that an orbit-averaged description applies.  The second requires that a typical noise force not appreciably tilt the barrier. Using~Eq.~\eqref{eq:chi_exact}, the condition $\chi \ll 1$ requires $\omega_c / \omega_0 \gg 8h$ for $\omega_c\ll \omega_0$ and $\alpha = 2$.  Thus, the weak-noise regime is a wedge in the $(h, \omega_c / \omega_0)$ plane rather than a single limit.

The limits $h\to0$ and $\omega_c/\omega_0\to0$ do not commute.  Taking $h\to 0$ at fixed $\omega_c/\omega_0$ sends $\chi\to 0$ and recovers pure gradual heating.  Taking $\omega_c\to 0$ at fixed $h$ instead sends $\chi\to\infty$ for $\alpha \geq 1$, and quasi-static barrier suppression dominates.  Having isolated the regime in which gradual heating dominates, we now calculate the escape time there.

\section{Stochastic averaging}\label{section:stochastic_averaging}

Noise-driven nonlinear oscillators are analytically challenging, in part due to the multi-dimensional boundary condition which determines the first-passage condition.  
% This makes the first-passage problem, even for the linear simple harmonic oscillators, subtle and analytically challenging~\cite{Hanggi1985, CRANDALL_1970}.  
In the weak-noise limit, averaging schemes~\cite{ROBERTS_1986} derive a first-order It\^o stochastic differential equation from second-order differential equations such as~Eq.~\eqref{eq:duffing}.  These techniques lead to effective rate-type equations for the trajectories.  While these approximations can introduce systematic errors, they offer quantitative and qualitative insight into the dynamics.

Averaging techniques use a transformation from $(x, \dot x)$, which both evolve on similar timescales, to a pair of variables where one of the variables evolves more slowly than the other.  The fast variable is then time-averaged, leaving the dynamics of the slow variable.  Many choices of transformations are possible.  We investigate two transformation choices: amplitude-phase variables, and action-angle variables.  The former has been treated for the finite-depth oscillator, is intuitive, and is simpler analytically.  However, these advantages come at the cost of predictive power.  While amplitude-phase averaging is a useful heuristic, it poses several challenges of interpretation.  Action-angle variables resolve many of these issues, but come at the cost of significant analytical complexity.

\subsection{Amplitude-phase averaging}

The natural first attempt is to use a standard transformation from position-velocity variables $(x, \dot{x})$ to amplitude-phase variables $(A,\phi)$, where $x(t) = A\cos[\omega_\text{eff}(A)t + \phi(t)],\; \dot x(t) = - \omega_\text{eff}(A) A \sin[\omega_\text{eff}(A)t + \phi(t)]$, and where $\omega^2_\text{eff}(A) = \omega_0^2(1-3A^2/4r^2)$ is an amplitude-dependent frequency obtained using harmonic balance.  This transformation has been used for the finite-depth oscillator~\cite{Kougioumtzoglou_Spanos_2014, Kougioumtzoglou_2016} and yields
\begin{equation}\label{eq:averaged_SDE}
dA =  \bigg[\frac{S_\xi\big(\omega_\text{eff}(A)\big)}{4Am^2\omega_\text{eff}^2(A)} \bigg]dt + \sqrt{\frac{S_\xi\big(\omega_\text{eff}(A)\big)}{2m^2\omega_\text{eff}^2(A)}}dW.
\end{equation}
We use it to find a closed-form approximation for the mean first-passage time for a particle starting at $A=0$ (which is an entrance boundary) to $A_\text{max} = 2r/\sqrt{3}$ (which is where the effective frequency vanishes).  We obtain
\begin{equation}\label{eq:ratio_digamma}
    T_1 = \frac{4m^2\omega_0^2r^2}{3 S_0(\omega_0^\alpha + \omega_c^\alpha)}\bigg[\omega_0^\alpha\frac{\psi(3+\alpha/2) + \gamma_E}{2+\alpha/2} + \frac{3\omega_c^\alpha}{4}\bigg],
\end{equation}
where the spectral exponent enters through the digamma function $\psi(x)$, and where $\gamma_E\approx0.57721$ is the Euler--Mascheroni constant.  

This result, however, contains $O(1)$ errors.  At $\alpha = 0$, it gives $T_1 = 4m\Delta / S_0$, which is twice the martingale result~Eq.~\eqref{eq:martingale_anchor}.  This deviation is primarily due to two approximations: the amplitude-dependent frequency is only controlled for $u \equiv 3A^2/4r^2 \ll 1$ whereas the first-passage problem integrates through $u=O(1)$, and the sinusoidal ansatz limits the cycle-averaged maximum energy to $\langle E\rangle_\text{max} = 8\Delta/9$ and hence has no exact analogue for the escape criterion.  We can therefore use~Eq.~\eqref{eq:ratio_digamma} as a heuristic indication of the \emph{shape} of the escape time dependence on $\omega_c/\omega_0$ and $\alpha$.  See Appendix~\ref{appendix:amplitude_phase_averaging} for derivations and a discussion of the approximations.  In order to avoid these systematic errors coming from uncontrolled approximations, we must instead use a transformation that treats the anharmonicity exactly.

\subsection{Action-angle averaging}
Amplitude-phase averaging is first-order in both the noise and the anharmonicity.  Furthermore, these variables only exist inside the sinusoidal ansatz, whereas for a softening oscillator the phase-space orbit is not an ellipse.  Under noise-free flow, $A$ wobbles within a single period.  By contrast, action-angle variables follow from a canonical phase-space transformation.  The action-angle averaging is therefore exact in the anharmonicity, but still first-order in the noise.

For an unperturbed Hamiltonian of the form $H_0(x,p) = p^2 / 2m + V(x)$, the action-angle transformation $(x,p)\to(I,\phi)$ is defined as~\cite{Mallick_2009}
\begin{equation}
    I(E) = \frac{1}{2\pi}\oint p\,dx,
\end{equation}
and
\begin{equation}
    \phi = \Omega(E)\int_0^x \frac{dy}{\sqrt{2[E-V(y)]/m}},
\end{equation}
with
\begin{equation}\label{eq:non_linear_frequency_def}
    \Omega(E) = \Bigg(\frac{2}{\pi}\int_0^{a(E)} \frac{dy}{\sqrt{2[E-V(y)]/m}}\Bigg)^{-1},
\end{equation}
where $E=V(a)$.  For every bound orbit the action then satisfies
\begin{equation}\label{eq:I_to_E}
    E = H(I), \quad \frac{dI}{dE}  = \frac{1}{\Omega(I)}.
\end{equation}
In the absence of external noise, the action is a constant of motion.  Noise breaks the invariance and couples to the action
\begin{equation}
    \frac{dI}{dt} = \xi(t) \frac{\partial x(I,\phi)}{\partial\phi}.
\end{equation}
This form also highlights connections to the interaction Hamiltonian $\hat H_\text{int} = -f(t)\hat x$ used in quantum treatments of ion heating~\cite{james1998_PRL}.  An effective Markovian description is then achieved through Fourier expansion of the exact nonlinear orbits, and under the weak-noise approximation ($T_1\gg1/\omega_c$, which for $\alpha\ge2$ follows from $\chi\ll1$).  The resulting Fokker--Planck equation for the action has a divergence form~\cite{Mallick_2009}
\begin{equation}\label{eq:diffusion_FPK}
    \frac{\partial P}{\partial t} = \frac{\partial}{\partial I} \bigg[ D(I) \frac{\partial P}{\partial I}\bigg],
\end{equation}
where 
\begin{equation}\label{eq:action_diffusion}
    D(I) = \sum_{n=1}^\infty n^2 |x_n(I)|^2 S_\xi[n\Omega(I)].
\end{equation}
We define $x(I,\phi) = \sum_{n=-\infty}^\infty x_n(I)e^{in\phi}$, where $x_{-n} = x_n^*$.  This shows that the noise enters the dynamics through the harmonics of the nonlinear periodic orbit.  The normalized escape time to the physical barrier $I = I(\Delta)$ is
\begin{equation}\label{eq:action_escape_time_complete}
    \tau_\text{esc} = \frac{S_0}{2m\Delta}\int_0^{I(\Delta)} \frac{I}{D(I)}dI.  
\end{equation}
We approximate $D(I)$ using Parseval's identity $\big(\sum_{n=1}^\infty n^2|x_n(I)|^2 = I/2m\Omega(I)\big)$, and using the spectral-flatness-over-harmonics approximation $S_\xi(n\Omega) \to S_\xi(\Omega)$ (which results in a lower bound on $T_1$ for decreasing spectra) yielding
\begin{align}\label{eq:action_diffusion_simplified}
    D(I) \simeq \frac{I}{2m\Omega(I)}S_\xi[\Omega(I)],
\end{align}
which is transformed back into an energy variable using~Eq.~\eqref{eq:I_to_E}.  Appendix~\ref{appendix:fundamental_harmonic} evaluates this approximation against the full harmonic sum, and Fig.~\ref{fig:fundamental_harmonic_approximation} shows its effect on the predicted escape time.  This yields
\begin{equation}\label{eq:elliptic_integral}
    \tau_\text{esc} \simeq\frac{S_0}{\Delta}\int_0^\Delta \frac{dE}{S_\xi[\Omega(E)]}. 
\end{equation}
The remaining nontrivial element is now the exact nonlinear frequency of the oscillator $\Omega(E)$~Eq.~\eqref{eq:non_linear_frequency_def}, which is a well-known result from analytical mechanics~\cite{Gendelman_2024}
\begin{equation}
    \Omega\bigg(\frac{E}{\Delta}\bigg) = \frac{\pi\omega_0}{2\mathcal{K}(k)}\sqrt{\frac{1+\sqrt{1-\frac{E}{\Delta}}}{2}}, \, k^2 = \frac{1-\sqrt{1-\frac{E}{\Delta}}}{1+\sqrt{1-\frac{E}{\Delta}}}
\end{equation}
where $\mathcal{K}$ is the complete elliptic integral of the first kind.  At small amplitudes, $\Omega^2(a) = \omega_0^2\big(1-3a^2 / 4r^2 + \mathcal{O}(a^4 / r^4)\big) $, which reproduces the first-order amplitude-dependent frequency used in the previous section.  

In the white-noise case,~Eq.~\eqref{eq:elliptic_integral} gives $\tau_\text{esc} = 1$, which reproduces the martingale result.  Colored noise spectra~Eq.~\eqref{eq:power_law_definition_with_cutoff} in the limit $\omega_c\to 0$ become a pure power law and the integral simplifies to
\begin{equation}\label{eq:tau_pure_power}
\tau_{\rm esc}\simeq\int_0^1\left[\frac{\Omega(E)}{\omega_0}\right]^{\alpha}
d\!\left(E/\Delta\right).
\end{equation}
For $\alpha \gg 1$,~Eq.~\eqref{eq:tau_pure_power} gives $\tau_\text{esc} \simeq 16/ (3\alpha)$.  Also, for $\omega_c \ll \omega_0$, the orbit satisfies $\Omega(E) < \omega_c$ only in an exponentially narrow window near the boundary, because $\Omega(E)$ vanishes inversely logarithmically there.  The weak-noise theoretical prediction~Eq.~\eqref{eq:elliptic_integral} becomes nearly cutoff independent as $\omega_c\to 0$, despite the total force variance growing without bound as $\omega_c\to0$ at fixed $S_\xi(\omega_0)$.  This eventually violates the $\chi\ll 1$ condition for weak noise, showing the importance of the ordered limits discussed above.  For $\alpha\ge 2$, this condition also implies that the noise correlation time should be short compared to the gradual escape time, even if it is long relative to one orbit.    

The integral~Eq.~\eqref{eq:elliptic_integral} has no closed form for general spectra of the form~Eq.~\eqref{eq:power_law_definition_with_cutoff} (and likewise the integral~Eq.~\eqref{eq:tau_pure_power}), so we use numerical integration to calculate $\tau_\text{esc}$, and compare with simulations below.  

We note that~Eqs.~\eqref{eq:elliptic_integral}~and~\eqref{eq:tau_pure_power} generalize easily to other potential types, such as Gaussian potentials, which are relevant to tweezer experiments.

\begin{figure}
    \centering
    \includegraphics[width=1\linewidth]{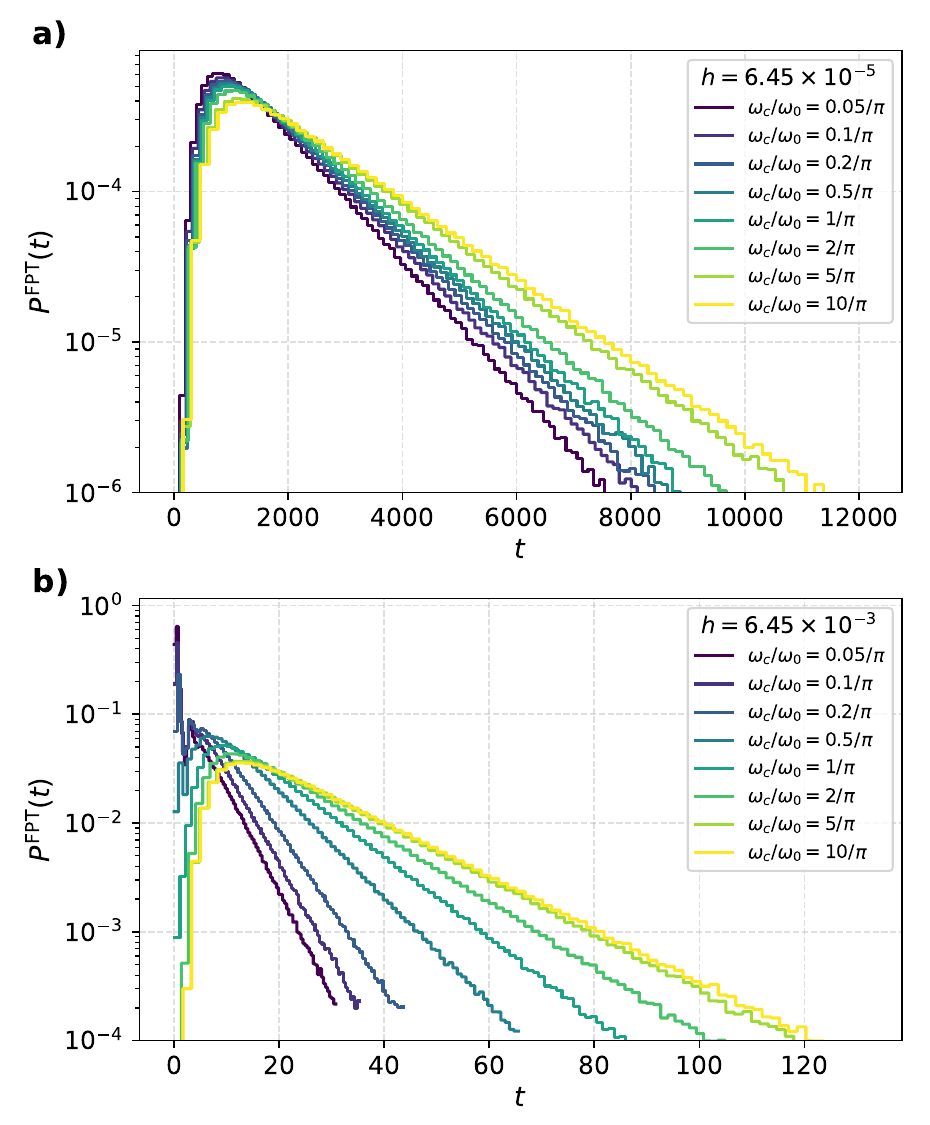}
    \caption{$P^\text{FPT}$ of the O-U-driven finite-depth oscillator~Eq.~\eqref{eq:duffing} with $m=1$ for (a) $h = 6.45\times 10^{-5}$, where $\chi=0.02$--$0.18$ for the cutoffs shown.  Here, changing $\omega_c$ has minimal effect on the shape of $P^\text{FPT}$; and (b) $h = 6.45\times 10^{-3}$, where $\chi=0.19$--$1.8$ for the cutoffs shown.  There, one can see that changing $\omega_c$ not only strongly modifies the long-time exponential tail of $P^\text{FPT}$, but there is also a qualitative change in the short-time behavior.}
    \label{fig:fptd_2_epsilons}
\end{figure}
\begin{figure}
    \centering
    \includegraphics[width=1\linewidth]{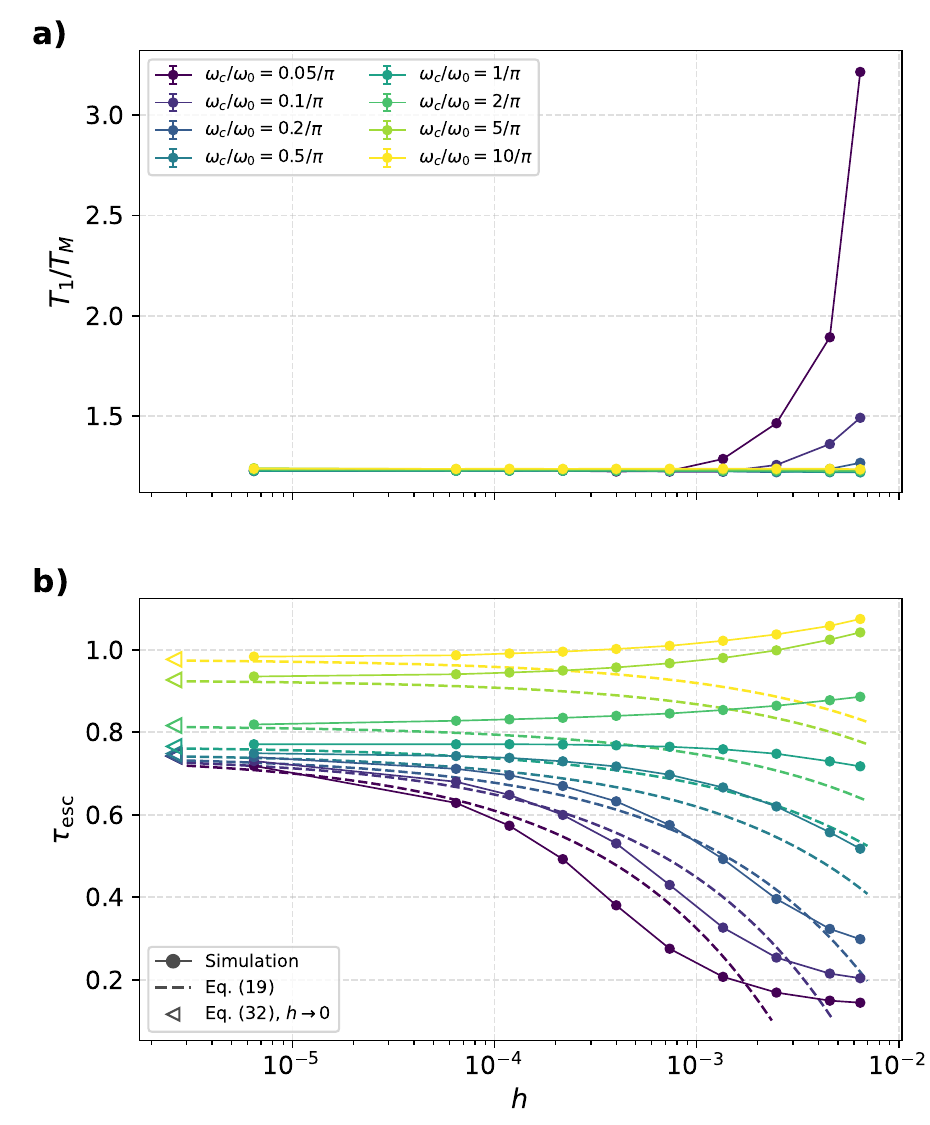}
    \caption{Escape statistics of the O-U-driven finite-depth oscillator with $m=1$, $\omega_0=\pi$, and $S_\xi(\omega_0)=2$, for various cutoffs $\omega_c/\omega_0$. (a) Ratio of the mean to the median first-passage time, $T_1/T_M$. (b) Normalized escape time $\tau_{\rm esc}$. Points connected by solid lines are simulations.  Triangles mark the weak-noise prediction~Eq.~\eqref{eq:elliptic_integral}, which is independent of $h$ and is drawn to the left of the simulated range.  Dashed lines include the quasi-static correction~Eq.~\eqref{eq:tilt_correction}.  They follow the simulations where $\chi\ll1$ and $\omega_c\ll\omega_0$ and depart from them outside this range.  Error bars are $1\sigma$ sampling errors and are mostly smaller than the markers.  Simulations sweep $h$ by varying $\Delta$.}
    \label{fig:ratio}
\end{figure}

\section{Simulation results}
\subsection{O-U noise}
We first simulate the finite-depth oscillator~Eq.~\eqref{eq:duffing} driven by O-U noise where $\xi(t)$ evolves according to~Eq.~\eqref{eq:OU_SDE}.  See Appendix~\ref{appendix:numerical_simulation_detail} for simulation details.  We isolate the effect of spectral color by holding the noise power $S_0$ at $\omega_0$ constant.  The particle begins at the bottom of the well, $x(0)=v(0)=0$, and the initial O-U force is sampled from its stationary Gaussian distribution.  For slowly varying noise, the force is approximately static during the initial oscillation, but is already present when the particle is released at rest at the untilted trap center.  Appendix~\ref{appendix:initial_state_prep} compares this choice of initial preparation with a particle cooled into the tilted well.  The two preparations agree throughout the weak-noise regime, and differ only when $\chi$ is not small.

Figure~\ref{fig:fptd_2_epsilons} shows numerically evaluated first-passage-time distributions for two values of $h$.  For the smaller value, $h=6.45\times10^{-5}$, the particle completes many ($\sim 10^3$) oscillations before escape.  Changing $\omega_c$ within the range shown modifies the escape time but produces comparatively small changes in the shape of the distribution.  For the larger value, $h=6.45\times10^{-3}$, the particle completes few oscillations before escaping ($\approx 12$ in the white-noise case) and, for small $\omega_c/\omega_0$, the distributions are sharply peaked at short times.  This peak comes from realizations in which the slowly varying force is already large when the particle is released.  For sufficiently large force, the particle starts above the lowered saddle, or the tilted well has no minimum at all, and the particle leaves on this first oscillation.  A particle cooled into the tilted well before release receives no such kick (Appendix~\ref{appendix:initial_state_prep}).

To quantify the change in distribution shape at fixed $h$ as we vary $\omega_c/\omega_0$, we show the ratio of the mean to the median first-passage time $T_1/T_M$~\cite{Teitsworth_2012} in Fig.~\ref{fig:ratio}(a).  For small $h$,  $T_1/T_M$ is nearly independent of $\omega_c$ within the range shown, and approaches a constant, indicating that changes in the spectrum primarily rescale the escape time while leaving the shape of the distribution unchanged.  As $h$ (and therefore $\chi$) increases, $T_1/T_M$ becomes cutoff-dependent, signaling the appearance of a distinct short-time escape mechanism.

% At long times, the first-passage-time distributions decay exponentially.  The long-time decay rate depends on $\omega_c$ when $\omega_c/\omega_0\lesssim1$, but becomes insensitive to $\omega_c$ in the limit $\omega_c/\omega_0\gg1$, where the spectrum is approximately white over the dynamically relevant frequency range.  

We compute the normalized escape time $\tau_{\rm esc}$ and show the results in Fig.~\ref{fig:ratio}(b).  At the smallest values of $h$ shown, the curves for sufficiently small $\omega_c/\omega_0$ converge.  In this regime, the action changes incrementally at each orbit, and escape occurs through gradual heating.  Equation~\eqref{eq:elliptic_integral} describes the normalized escape time in this regime.  As $h$ increases, the simulated values of $\tau_\text{esc}$ separate according to $\omega_c/\omega_0$, indicating the onset of additional low-frequency escape mechanisms such as barrier tilting.  

The dashed curves are the product of the weak-noise prediction~Eq.~\eqref{eq:elliptic_integral} and the tilt correction~Eq.~\eqref{eq:tilt_correction}.  They follow the simulations while $\chi\ll1$ and $\omega_c\ll\omega_0$, and separate from them where either condition fails: at large $\chi$ the barrier modulation is no longer perturbative, and for $\omega_c\gtrsim\omega_0$ the force is no longer quasi-static over an oscillation.  In the latter case $\tau_{\rm esc}$ instead rises slowly with $h$ and can exceed unity.

\begin{figure}
    \centering
    \includegraphics[width=1\linewidth]{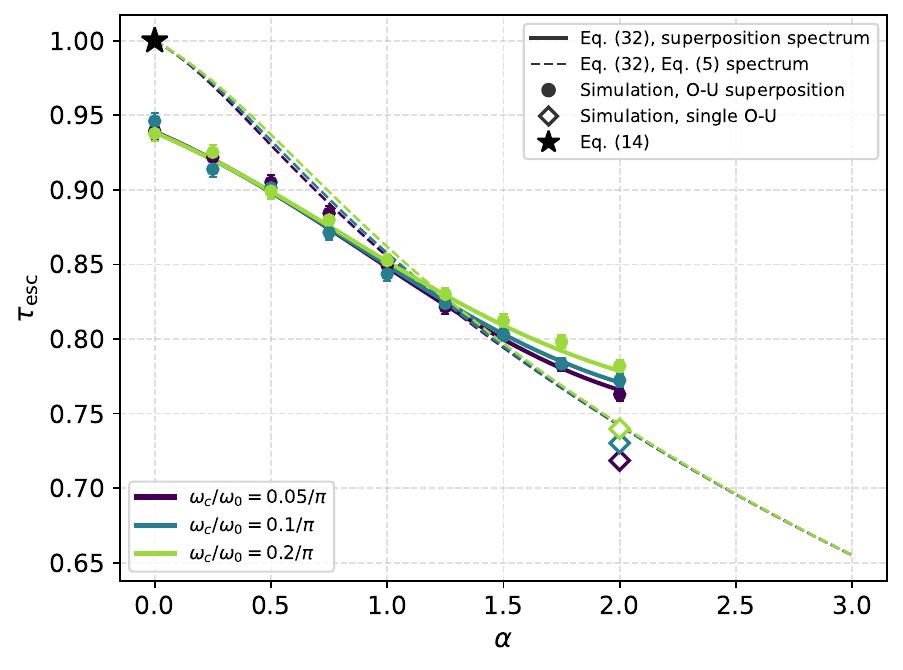}
    \caption{Normalized escape time $\tau_{\rm esc}=\dot E_{\rm cold}/\dot E_{\rm esc}$ for the finite-depth oscillator of~Eq.~\eqref{eq:duffing} with $h=6.45\times10^{-6}$, driven by power-law noise with low-frequency cutoff $\omega_c$ and fixed $S_0=2$, so that $\dot E_{\rm cold}=1$ throughout.  Filled circles are Monte Carlo first-passage times using the O-U superposition of Appendix~\ref{appendix:colored_noise_generation}.  Solid curves are the action-angle prediction~Eq.~\eqref{eq:elliptic_integral} evaluated with the exact spectrum of the $N=20$ superposition.  Dashed curves are the same prediction for the idealized spectrum~Eq.~\eqref{eq:power_law_definition_with_cutoff}. Both are independent of $h$. Open diamonds are single O-U processes, whose spectrum is exactly~Eq.~\eqref{eq:power_law_definition_with_cutoff} with $\alpha=2$.  They have $\chi=0.057$, $0.040$, and $0.028$ for $\omega_c/\omega_0=0.05/\pi$, $0.1/\pi$, and $0.2/\pi$, and are the leftmost points of Fig.~\ref{fig:ratio}(b). The star is the white-noise result~Eq.~\eqref{eq:martingale_anchor}. Error bars are $1\sigma$ sampling errors.}
    \label{fig:colored_alpha}
\end{figure}
\begin{figure}
    \centering
    \includegraphics[width=1\linewidth]{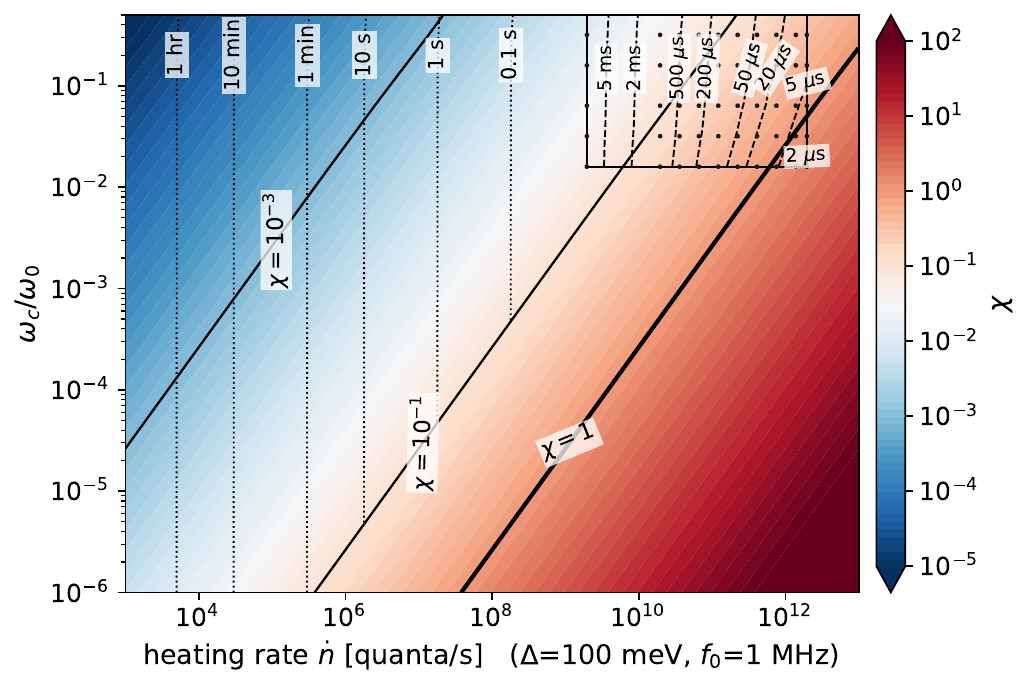}
    \caption{Escape-time regime map for a representative trapped-ion system with $\omega_0=2\pi\times 1$~MHz and $\Delta=100$~meV, driven by O-U noise ($\alpha=2$).  The heating rate $\dot{\bar n} $ and the trap depth $\Delta$ fix $h$ through~Eq.~\eqref{eq:h_definition}.  The color scale is the barrier-tilt parameter $\chi$ of~Eq.~\eqref{eq:chi_definition}.  Solid curves show contours of $\chi$.  Dashed contours inside the box are mean escape times interpolated from the simulation data of Fig.~\ref{fig:ratio}(b) (points shown), and dotted contours are the weak-noise escape times calculated from theory~Eq.~\eqref{eq:elliptic_integral} and are drawn only where $\chi<10^{-1}$, an approximate boundary for their region of applicability.  The theoretical contours are almost independent of the cutoff as expected.  Typical trapped-ion heating rates, $\dot {\bar n}\lesssim10^{4}$ quanta/s and trapping lifetimes of $\gg 1$~s, place most real ion traps several decades inside the weak-noise region.}
    \label{fig:heatmap}
\end{figure}

%############################
\subsection{Colored noise}\label{section:colored_noise_results}
%############################

We now simulate more general forcing spectra of the form~Eq.~\eqref{eq:power_law_definition_with_cutoff} acting on the finite-depth oscillator~Eq.~\eqref{eq:duffing} with $m=1$.  We approximate power-law spectra with exponent $0\leq\alpha\leq 2$ using a weighted superposition of O-U processes inspired by McWhorter's explanation of $1/f$ noise~\cite{Mcworther}.  See Appendix~\ref{appendix:colored_noise_generation} for details.

Since the weak-noise limit is more relevant to atom trap experiments, our simulations focus on that regime.  We choose $\Delta$ sufficiently large that the weak-noise criteria are satisfied throughout the parameter range shown.  This suppresses the direct quasi-static escape mechanism and barrier tilting effects observed in the O-U simulations and isolates the effect of spectral color on gradual heating and escape.  We show the numerically evaluated $\tau_\text{esc}$ for variable $\alpha$ in Fig.~\ref{fig:colored_alpha}.  The normalized escape time decreases monotonically with $\alpha$. This behavior follows from the softening of the trap: as the particle approaches the barrier, its oscillation frequency decreases, and a spectrum with larger $\alpha$ contains more power at the frequencies sampled late in the trajectory.

Figure~\ref{fig:colored_alpha} shows the weak-noise predictions of Eq.~\eqref{eq:elliptic_integral} for two different spectra: the ideal power-law spectrum (dashed lines), and the exact spectrum of the $N=20$ superposition (solid lines).  The two agree around $\alpha\sim 1$ because the superposition method is most faithful to Eq.~\eqref{eq:power_law_definition_with_cutoff} there [see Appendix~\ref{appendix:colored_noise_generation} and Fig.~\ref{fig:power_spectra}(c)].  The open diamonds shown in Fig.~\ref{fig:colored_alpha} are single O-U processes, whose spectra follow Eq.~\eqref{eq:power_law_definition_with_cutoff} with $\alpha = 2$.  These points are also the leftmost points of Fig.~\ref{fig:ratio}(b).  At matched $\omega_c/\omega_0$, these carry $2.6 - 3.7$ times the total force variance of the superposition method, and correspondingly have a larger tilt parameter.  They therefore fall below the superposition points, and approach the ideal-spectrum prediction as $\chi$ decreases, consistent with a residual quasi-static tilt.  The action-angle prediction reduces $\tau_\text{esc}$ from unity to about $0.59$ over $0\leq\alpha\leq4$ (the colored-noise simulations shown here cover $0\leq\alpha\leq2$).  We now investigate whether this can account for the trap lifetimes observed in Ref.~\cite{Stick2006}.

\section{Application to Experimental Systems}

Figure~\ref{fig:heatmap} places our results in experimental units for fixed trap frequency $\omega_0$ and depth $\Delta$, and assuming $\alpha=2$.  The color scale shows $\chi$ as a function of the measured heating rate $\dot{\bar n}$ and the cutoff ratio $\omega_c/\omega_0$.  The superposed contours show the corresponding mean escape time $T_1$.  When $\chi\ll1$, the lifetime is governed predominantly by gradual heating and is therefore nearly independent of $\omega_c$ at fixed $\dot{\bar n}$.  As $\chi$ approaches unity, the contours acquire a stronger cutoff dependence, indicating that quasi-static modification of the barrier can no longer be neglected.

We apply these results to investigate the experimental measurements made with a microfabricated ion trap~\cite{Stick2006}, which had a depth of $\Delta \sim 80$~meV, axial heating rates of $\dot{\bar n} \sim 10^6$~quanta/s at $\omega_0 \sim 2\pi\times 0.9$~MHz, and reported lifetimes of $T_1\sim 0.1$~s without laser cooling.   In this example, $\Delta / \hbar\omega_0 \approx 2.4\times 10^{7}$, so the motion is deep in the classical regime and a classical treatment is appropriate.  We use the measured axial heating rate and reported trap depth as representative parameters of the one-dimensional model.  The experiment does not determine the escape direction.  We extract $h\sim 3\times 10^{-9}$, $\tau_\text{esc} \sim 0.004$, and $r\sim 94~\mu$m.  Ref.~\cite{Stick2006} estimates $\Delta/T_1$ to be about 100 times the measured heating rate, i.e.,\ $\tau_\text{esc}\sim0.01$.  Within the gradual-heating regime, Eq.~\eqref{eq:tau_pure_power} reaches $\tau_{\rm esc} \simeq0.004$ only for an extremely steep spectrum, $\alpha\sim10^3$, and $\tau_{\rm esc}\simeq0.01$ only for $\alpha\approx500$.  This is far outside the range of exponents normally observed in trapped-ion electric-field noise.  Thus, within the present model, experimentally plausible power-law spectral color cannot by itself account for the reported lifetime.

Substantial quasi-static barrier modification is not excluded by the available measurements.  An order-unity tilt parameter $\chi\sim1$ corresponds to an rms electric field satisfying $eE_{\rm rms}r\sim\Delta$, giving $E_{\rm rms}\sim1\times10^3$~V/m, or an electric-potential variation $E_{\rm rms}r\sim100$~mV across the characteristic trap dimension.  In the harmonic approximation, such a field would displace the equilibrium position by approximately $r/4$.  It should therefore produce observable displacement or excess micromotion and could, in principle, be constrained by ion-position and micromotion-compensation measurements.  Ref.~\cite{Stick2006} discusses accumulated charge on insulating surfaces and contact potentials as possible sources of large stray fields.

This estimate does not establish barrier tilting as the explanation for the observed loss.  The regime $\chi\sim1$ lies outside both the perturbative correction of Eq.~\eqref{eq:tilt_correction} and the parameter range over which our simulations quantitatively determine the lifetime.  Moreover, for an O-U spectrum normalized to the measured resonant noise, achieving $\chi\sim1$ requires
$\omega_c\approx2\pi\times0.03$~Hz, corresponding to a correlation time
$\omega_c^{-1}\approx5$~s, much longer than the reported $0.1$-s lifetime.  The force would therefore be effectively static during an individual experimental realization, and the inferred loss would depend on the preparation protocol and on the distribution of slowly varying fields between realizations.  

Many additional escape mechanisms lie outside the presented model, such as background gas collisions, narrow mechanical or electrical resonances, and time-dependent patch or charging fields.  Furthermore, we have not considered escape along the transverse trap axes, which suffer from parametric rf heating and have independent pseudopotential depths and heating rates.  Consequently, our comparison supports only the narrower conclusion that ordinary power-law spectral color cannot explain the reported lifetime through gradual axial heating alone.  It does not identify the mechanism responsible for the observed ion loss.

\section{Conclusion}
We have investigated the escape dynamics of particles confined in finite-depth traps subject to power-law noise spectra.  We have approached this problem using both numerical and theoretical tools.  We identify two distinct dynamical regimes: the weak- and strong-noise limits.  The weak-noise case applies when the trap depth reduction parameter from barrier tilting $\chi\ll 1$, which implies the fraction of trap depth acquired per oscillation $4\pi h\ll 1$.  In the weak-noise limit, we calculate the normalized escape time $\tau_\text{esc} = \dot{E}_{\mathrm{cold}} / \dot{E}_{\mathrm{esc}}$ using stochastic averaging techniques.  We find that in this limit, increasing $\alpha$ only yields a modest reduction in $\tau_\text{esc}$, and a far less dramatic effect than one could have inferred from the experiment of Ref.~\cite{Stick2006}.  This suggests that either this experiment suffered from substantial barrier tilting, or that there were other unrelated escape mechanisms such as parametric rf heating~\cite{Wineland1998, vanMourik_2022}, or that escape proceeded along a transverse direction, for which no heating rate was measured.  

The theoretical and simulation techniques developed here should guide future experiments, particularly those aiming to mitigate or eliminate high-$\alpha$ noise from physical systems~\cite{Mehmandoost2024}.  Future work on the theoretical side should explore the approximate behavior near the boundary, as well as the intermediate- and strong-noise regimes, where the averaging method discussed above fails.  It will also be important to test the predictions against alternative trap potentials and independently measured low- and high-frequency spectral cutoffs.

\vspace{1cm}
\textbf{Acknowledgments} This work is supported by the DOE Quantum Systems Accelerator (DE-FOA-0002253) and the NSF STAQ Program (PHY-1818914).  This work was also funded by NSF under QLCI: Center for Robust Quantum Simulation OMA-2120757.  J.M.R. was supported
by the Goshaw Fellowship.

\bibliography{refs}

@article{Barkai_2013,
  title = {Fluctuations of $1/f$ Noise and the Low-Frequency Cutoff Paradox},
  author = {Niemann, Markus and Kantz, Holger and Barkai, Eli},
  journal = {Phys. Rev. Lett.},
  volume = {110},
  issue = {14},
  pages = {140603},
  numpages = {5},
  year = {2013},
  month = {Apr},
  publisher = {American Physical Society},
  doi = {10.1103/PhysRevLett.110.140603},
  url = {https://link.aps.org/doi/10.1103/PhysRevLett.110.140603}
}

@Article{Bonus2025,
author={Bonus, F.
and Knapp, C.
and Valahu, C. H.
and Mironiuc, M.
and Weidt, S.
and Hensinger, W. K.},
title={Ultrasensitive single-ion electrometry in a magnetic field gradient},
journal={Nature Physics},
year={2025},
month={Aug},
day={01},
volume={21},
number={8},
pages={1189-1195},
issn={1745-2481},
doi={10.1038/s41567-025-02887-9},
url={https://doi.org/10.1038/s41567-025-02887-9}
}

@article{Brownnutt2015,
  title = {Ion-trap measurements of electric-field noise near surfaces},
  author = {Brownnutt, M. and Kumph, M. and Rabl, P. and Blatt, R.},
  journal = {Rev. Mod. Phys.},
  volume = {87},
  issue = {4},
  pages = {1419--1482},
  numpages = {64},
  year = {2015},
  month = {Dec},
  publisher = {American Physical Society},
  doi = {10.1103/RevModPhys.87.1419},
  url = {https://link.aps.org/doi/10.1103/RevModPhys.87.1419}
}

@article{CRANDALL_1980,
title = {Non-gaussian closure for random vibration of non-linear oscillators},
journal = {International Journal of Non-Linear Mechanics},
volume = {15},
number = {4},
pages = {303-313},
year = {1980},
issn = {0020-7462},
doi = {https://doi.org/10.1016/0020-7462(80)90015-3},
url = {https://www.sciencedirect.com/science/article/pii/0020746280900153},
author = {Stephen H. Crandall}
}

@article{Dutta_1981,
  title = {Low-frequency fluctuations in solids: $\frac{1}{f}$ noise},
  author = {Dutta, P. and Horn, P. M.},
  journal = {Rev. Mod. Phys.},
  volume = {53},
  issue = {3},
  pages = {497--516},
  numpages = {0},
  year = {1981},
  month = {Jul},
  publisher = {American Physical Society},
  doi = {10.1103/RevModPhys.53.497},
  url = {https://link.aps.org/doi/10.1103/RevModPhys.53.497}
}

@book{Gardiner2009,
    edition={4th ed.}, 
    series={Springer series in synergetics}, title={Stochastic methods: a handbook for the natural and social sciences}, publisher={Springer}, author={Gardiner, C. W.}, year={2009}, collection={Springer series in synergetics} }

@article{Gibbons_2011,
  title = {Nondestructive Fluorescent State Detection of Single Neutral Atom Qubits},
  author = {Gibbons, Michael J. and Hamley, Christopher D. and Shih, Chung-Yu and Chapman, Michael S.},
  journal = {Phys. Rev. Lett.},
  volume = {106},
  issue = {13},
  pages = {133002},
  numpages = {4},
  year = {2011},
  month = {Mar},
  publisher = {American Physical Society},
  doi = {10.1103/PhysRevLett.106.133002},
  url = {https://link.aps.org/doi/10.1103/PhysRevLett.106.133002}
}

@article{Raychaudhuri2002,
title = {Measurement of 1/f noise and its application in materials science},
journal = {Current Opinion in Solid State and Materials Science},
volume = {6},
number = {1},
pages = {67-85},
year = {2002},
issn = {1359-0286},
doi = {https://doi.org/10.1016/S1359-0286(02)00025-6},
url = {https://www.sciencedirect.com/science/article/pii/S1359028602000256},
author = {A.K. Raychaudhuri}
}

@book{feller_boundary,
    author = {Feller, William},
    title = {An Introduction to Probability Theory and Its
Applications},
    volume = {2},
    edition = {3},
    publisher  = {Wiley},
    year = {1971}
}

@article{Gendelman_2024,
author = {Genda, Attila and Fidlin, Alexander and Gendelman, Oleg},
title = {An alternative approach to averaging in nonlinear systems using classical probability density},
journal = {ZAMM - Journal of Applied Mathematics and Mechanics / Zeitschrift für Angewandte Mathematik und Mechanik},
volume = {104},
number = {6},
pages = {e202300432},
doi = {https://doi.org/10.1002/zamm.202300432},
url = {https://onlinelibrary.wiley.com/doi/abs/10.1002/zamm.202300432},
eprint = {https://onlinelibrary.wiley.com/doi/pdf/10.1002/zamm.202300432},
year = {2024}
}

@article{Grimm_2000,
title = {Optical Dipole Traps for Neutral Atoms},
journal = {Advances In Atomic, Molecular, and Optical Physics},
publisher = {Academic Press},
volume = {42},
pages = {95-170},
year = {2000},
doi = {https://doi.org/10.1016/S1049-250X(08)60186-X},
url = {https://www.sciencedirect.com/science/article/pii/S1049250X0860186X},
author = {Rudolf Grimm and Matthias Weidemüller and Yurii B. Ovchinnikov}
}

@article{Hanggi1990,
  title = {Reaction-rate theory: fifty years after Kramers},
  author = {H\"anggi, Peter and Talkner, Peter and Borkovec, Michal},
  journal = {Rev. Mod. Phys.},
  volume = {62},
  issue = {2},
  pages = {251--341},
  numpages = {0},
  year = {1990},
  month = {Apr},
  publisher = {American Physical Society},
  doi = {10.1103/RevModPhys.62.251},
  url = {https://link.aps.org/doi/10.1103/RevModPhys.62.251}
}

@article{Home_2011,
doi = {10.1088/1367-2630/13/7/073026},
url = {https://doi.org/10.1088/1367-2630/13/7/073026},
year = {2011},
month = {jul},
publisher = {},
volume = {13},
number = {7},
pages = {073026},
author = {Home, J P and Hanneke, D and Jost, J D and Leibfried, D and Wineland, D J},
title = {Normal modes of trapped ions in the presence of anharmonic trap potentials},
journal = {New Journal of Physics}
}

@article{Julia_rackauckas2017differentialequations,
  title={Differential{E}quations.jl--a performant and feature-rich ecosystem for solving differential equations in {J}ulia},
  author={Rackauckas, Christopher and Nie, Qing},
  journal={Journal of Open Research Software},
  volume={5},
  number={1},
  year={2017},
  publisher={Ubiquity Press}
}

@article{Kougioumtzoglou_2016,
author = {Ioannis A. Kougioumtzoglou  and Yuanjin Zhang  and Michael Beer },
title = {Softening Duffing Oscillator Reliability Assessment Subject to Evolutionary Stochastic Excitation},
journal = {ASCE-ASME Journal of Risk and Uncertainty in Engineering Systems, Part A: Civil Engineering},
volume = {2},
number = {2},
pages = {C4015001},
year = {2016},
doi = {10.1061/AJRUA6.0000828},
URL = {https://ascelibrary.org/doi/abs/10.1061/AJRUA6.0000828},
eprint = {https://ascelibrary.org/doi/pdf/10.1061/AJRUA6.0000828}
}

@article{Mallick_2002,
  title = {Anomalous diffusion in nonlinear oscillators with multiplicative noise},
  author = {Mallick, Kirone and Marcq, Philippe},
  journal = {Phys. Rev. E},
  volume = {66},
  issue = {4},
  pages = {041113},
  numpages = {14},
  year = {2002},
  month = {Oct},
  publisher = {American Physical Society},
  doi = {10.1103/PhysRevE.66.041113},
  url = {https://link.aps.org/doi/10.1103/PhysRevE.66.041113}
}

@article{Mallick_2004,
doi = {10.1088/0305-4470/37/17/008},
url = {https://doi.org/10.1088/0305-4470/37/17/008},
year = {2004},
month = {apr},
publisher = {},
volume = {37},
number = {17},
pages = {4769},
author = {Kirone Mallick and Philippe Marcq},
title = {On the stochastic pendulum with Ornstein–Uhlenbeck noise},
journal = {Journal of Physics A: Mathematical and General}
}

@article{Martin_2011,
  title = {First-passage and first-exit times of a Bessel-like stochastic process},
  author = {Martin, Edgar and Behn, Ulrich and Germano, Guido},
  journal = {Phys. Rev. E},
  volume = {83},
  issue = {5},
  pages = {051115},
  numpages = {16},
  year = {2011},
  month = {May},
  publisher = {American Physical Society},
  doi = {10.1103/PhysRevE.83.051115},
  url = {https://link.aps.org/doi/10.1103/PhysRevE.83.051115}
}

@book{Mcworther,
  author       = {A. L. McWhorter},
  title        = {1/f Noise and Germanium Surface Properties},
  booktitle    = {Semiconductor Surface Physics},
  editor       = {R. H. Kingston},
  publisher    = {University of Pennsylvania Press},
  address      = {Philadelphia},
  year         = {1957},
  pages        = {207--228}
}

@article{Mehmandoost2024,
  title = {Decoherence induced by a sparse bath of two-level fluctuators: Peculiar features of $1/f$ noise in high-quality qubits},
  author = {Mehmandoost, M. and Dobrovitski, V. V.},
  journal = {Phys. Rev. Res.},
  volume = {6},
  issue = {3},
  pages = {033175},
  numpages = {18},
  year = {2024},
  month = {Aug},
  publisher = {American Physical Society},
  doi = {10.1103/PhysRevResearch.6.033175},
  url = {https://link.aps.org/doi/10.1103/PhysRevResearch.6.033175}
}

@article{Mallick_2009,
  title = {Influence of the noise spectrum on the anomalous diffusion in a stochastic system},
  author = {Mallick, K.},
  journal = {Phys. Rev. E},
  volume = {80},
  issue = {1},
  pages = {011124},
  numpages = {10},
  year = {2009},
  month = {Jul},
  publisher = {American Physical Society},
  doi = {10.1103/PhysRevE.80.011124},
  url = {https://link.aps.org/doi/10.1103/PhysRevE.80.011124}
}

@article{Ornstein_1930,
  title = {On the Theory of the Brownian Motion},
  author = {Uhlenbeck, G. E. and Ornstein, L. S.},
  journal = {Phys. Rev.},
  volume = {36},
  issue = {5},
  pages = {823--841},
  numpages = {0},
  year = {1930},
  month = {Sep},
  publisher = {American Physical Society},
  doi = {10.1103/PhysRev.36.823},
  url = {https://link.aps.org/doi/10.1103/PhysRev.36.823}
}

@article{PLASZCZYNSKI2007,
author = {Plaszczynski, S.},
title = {Generating long streams of 1/f^α noise},
journal = {Fluctuation and Noise Letters},
volume = {07},
number = {01},
pages = {R1-R13},
year = {2007},
doi = {10.1142/S0219477507003635},
URL = {https://doi.org/10.1142/S0219477507003635},
eprint = {https://doi.org/10.1142/S0219477507003635}
}

@article{ROBERTS_1986,
title = {Stochastic averaging: An approximate method of solving random vibration problems},
journal = {International Journal of Non-Linear Mechanics},
volume = {21},
number = {2},
pages = {111-134},
year = {1986},
issn = {0020-7462},
doi = {https://doi.org/10.1016/0020-7462(86)90025-9},
url = {https://www.sciencedirect.com/science/article/pii/0020746286900259},
author = {J.B. Roberts and P.D. Spanos}
}

@article{Rossler_2010,
author = {R\"{o}\ss{}ler, Andreas},
title = {Runge–Kutta Methods for the Strong Approximation of Solutions of Stochastic Differential Equations},
journal = {SIAM Journal on Numerical Analysis},
volume = {48},
number = {3},
pages = {922-952},
year = {2010},
doi = {10.1137/09076636X},
URL = {https://doi.org/10.1137/09076636X},
eprint = {https://doi.org/10.1137/09076636X}
}

@article{Roldan_2023,
author = {E. Roldán and I. Neri and R. Chetrite and S. Gupta and S. Pigolotti and F. Jülicher and K. Sekimoto},
title = {Martingales for physicists: a treatise on stochastic thermodynamics and beyond},
journal = {Advances in Physics},
volume = {72},
number = {1-2},
pages = {1--258},
year = {2023},
publisher = {Taylor \& Francis},
doi = {10.1080/00018732.2024.2317494},
URL = {https://doi.org/10.1080/00018732.2024.2317494},
eprint = {https://doi.org/10.1080/00018732.2024.2317494}}

@book{Redner_2001, 
place={Cambridge}, title={A Guide to First-Passage Processes}, publisher={Cambridge University Press}, author={Redner, Sidney}, year={2001}}

@article{Kougioumtzoglou_Spanos_2014,
title = {Stochastic response analysis of the softening Duffing oscillator and ship capsizing probability determination via a numerical path integral approach},
journal = {Probabilistic Engineering Mechanics},
volume = {35},
pages = {67-74},
year = {2014},
note = {Stochastic Mechanics (SM12)},
issn = {0266-8920},
doi = {https://doi.org/10.1016/j.probengmech.2013.06.001},
url = {https://www.sciencedirect.com/science/article/pii/S0266892013000544},
author = {Ioannis A. Kougioumtzoglou and Pol D. Spanos}
}

@Article{Stick2006,
author={Stick, D.
and Hensinger, W. K.
and Olmschenk, S.
and Madsen, M. J.
and Schwab, K.
and Monroe, C.},
title={Ion trap in a semiconductor chip},
journal={Nature Physics},
year={2006},
month={Jan},
day={01},
volume={2},
number={1},
pages={36-39},
issn={1745-2481},
doi={10.1038/nphys171},
url={https://doi.org/10.1038/nphys171}
}

@article{Teitsworth_2012,
  title = {Noise-Induced Current Switching in Semiconductor Superlattices: Observation of Nonexponential Kinetics in a High-Dimensional System},
  author = {Bomze, Yu. and Hey, R. and Grahn, H. T. and Teitsworth, S. W.},
  journal = {Phys. Rev. Lett.},
  volume = {109},
  issue = {2},
  pages = {026801},
  numpages = {4},
  year = {2012},
  month = {Jul},
  publisher = {American Physical Society},
  doi = {10.1103/PhysRevLett.109.026801},
  url = {https://link.aps.org/doi/10.1103/PhysRevLett.109.026801}
}

@article{VANVINCKENROYE2018,
title = {Second-order moment of the first passage time of a quasi-Hamiltonian oscillator with stochastic parametric and forcing excitations},
journal = {Journal of Sound and Vibration},
volume = {427},
pages = {178-187},
year = {2018},
issn = {0022-460X},
doi = {https://doi.org/10.1016/j.jsv.2018.03.001},
url = {https://www.sciencedirect.com/science/article/pii/S0022460X18301627},
author = {H. Vanvinckenroye and V. Denoël}
}

@Article{Wong-Campos2016,
author={Wong-Campos, J. D.
and Johnson, K. G.
and Neyenhuis, B.
and Mizrahi, J.
and Monroe, C.},
title={High-resolution adaptive imaging of a single atom},
journal={Nature Photonics},
year={2016},
month={Sep},
day={01},
volume={10},
number={9},
pages={606-610},
issn={1749-4893},
doi={10.1038/nphoton.2016.136},
url={https://doi.org/10.1038/nphoton.2016.136}
}

@article{james1998_PRL,
  title = {Theory of Heating of the Quantum Ground State of Trapped Ions},
  author = {James, Daniel F. V.},
  journal = {Phys. Rev. Lett.},
  volume = {81},
  issue = {2},
  pages = {317--320},
  numpages = {0},
  year = {1998},
  month = {Jul},
  publisher = {American Physical Society},
  doi = {10.1103/PhysRevLett.81.317},
  url = {https://link.aps.org/doi/10.1103/PhysRevLett.81.317}
}

@article{Paladino2014,
  title = {$\mathbsf{1}/\mathbsfit{f}$ noise: Implications for solid-state quantum information},
  author = {Paladino, E. and Galperin, Y. M. and Falci, G. and Altshuler, B. L.},
  journal = {Rev. Mod. Phys.},
  volume = {86},
  issue = {2},
  pages = {361--418},
  numpages = {58},
  year = {2014},
  month = {Apr},
  publisher = {American Physical Society},
  doi = {10.1103/RevModPhys.86.361},
  url = {https://link.aps.org/doi/10.1103/RevModPhys.86.361}
}

@article{Turchette2000,
  title = {Heating of trapped ions from the quantum ground state},
  author = {Turchette, Q. A. and Kielpinski, D. and King, B. E. and Leibfried, D. and Meekhof, D. M. and Myatt, C. J. and Rowe, M. A. and Sackett, C. A. and Wood, C. S. and Itano, W. M. and Monroe, C. and Wineland, D. J.},
  journal = {Phys. Rev. A},
  volume = {61},
  issue = {6},
  pages = {063418},
  numpages = {8},
  year = {2000},
  month = {May},
  publisher = {American Physical Society},
  doi = {10.1103/PhysRevA.61.063418},
  url = {https://link.aps.org/doi/10.1103/PhysRevA.61.063418}
}

@article{Savard1997,
  title = {Laser-noise-induced heating in far-off resonance optical traps},
  author = {Savard, T. A. and O'Hara, K. M. and Thomas, J. E.},
  journal = {Phys. Rev. A},
  volume = {56},
  issue = {2},
  pages = {R1095(R)--R1098(R)},
  numpages = {0},
  year = {1997},
  month = {Aug},
  publisher = {American Physical Society},
  doi = {10.1103/PhysRevA.56.R1095},
  url = {https://link.aps.org/doi/10.1103/PhysRevA.56.R1095}
}

@article{vanMourik_2022,
  title = {rf-induced heating dynamics of noncrystallized trapped ions},
  author = {van Mourik, Martin W. and Hrmo, Pavel and Gerster, Lukas and Wilhelm, Benjamin and Blatt, Rainer and Schindler, Philipp and Monz, Thomas},
  journal = {Phys. Rev. A},
  volume = {105},
  issue = {3},
  pages = {033101},
  numpages = {17},
  year = {2022},
  month = {Mar},
  publisher = {American Physical Society},
  doi = {10.1103/PhysRevA.105.033101},
  url = {https://link.aps.org/doi/10.1103/PhysRevA.105.033101}
}

@article{Wineland1998,
  author  = {Wineland, D. J. and Monroe, C. and Itano, W. M. and
             Leibfried, D. and King, B. E. and Meekhof, D. M.},
  title   = {Experimental Issues in Coherent Quantum-State Manipulation
             of Trapped Atomic Ions},
  journal = {Journal of Research of the National Institute of Standards
             and Technology},
  volume  = {103},
  pages   = {259--328},
  year    = {1998},
  doi     = {10.6028/jres.103.019}
}

@article{Gehm_1998,
  title = {Dynamics of noise-induced heating in atom traps},
  author = {Gehm, M. E. and O'Hara, K. M. and Savard, T. A. and Thomas, J. E.},
  journal = {Phys. Rev. A},
  volume = {58},
  issue = {5},
  pages = {3914--3921},
  numpages = {0},
  year = {1998},
  month = {Nov},
  publisher = {American Physical Society},
  doi = {10.1103/PhysRevA.58.3914},
  url = {https://link.aps.org/doi/10.1103/PhysRevA.58.3914}
}

@article{Hanggi_colored_1988,
  title = {Dynamical systems: A unified colored-noise approximation},
  author = {Jung, Peter and H\"anggi, Peter},
  journal = {Phys. Rev. A},
  volume = {35},
  issue = {10},
  pages = {4464(R)--4466(R)},
  numpages = {0},
  year = {1987},
  month = {May},
  publisher = {American Physical Society},
  doi = {10.1103/PhysRevA.35.4464},
  url = {https://link.aps.org/doi/10.1103/PhysRevA.35.4464}
}

@article{Dykman_1997,
  title = {Quasimonochromatic noise: New features of fluctuations in noise-driven nonlinear systems},
  author = {Dykman, M. I. and McClintock, P. V. E. and Stein, N. D. and Stocks, N. G.},
  journal = {Phys. Rev. Lett.},
  volume = {67},
  issue = {8},
  pages = {933--936},
  numpages = {0},
  year = {1991},
  month = {Aug},
  publisher = {American Physical Society},
  doi = {10.1103/PhysRevLett.67.933},
  url = {https://link.aps.org/doi/10.1103/PhysRevLett.67.933}
}

\appendix

\clearpage
\widetext

\emph{Throughout the appendices we set $m = 1$, so that $\xi$ is an acceleration and $S_\xi$ an acceleration PSD.  The main-text results are recovered by $S_\xi\to S_\xi/m^2$ and $\Delta\to\Delta / m$.  }

\section{O-U-driven simple harmonic oscillator}\label{appendix:ou_sho}
Following~Eq.~\eqref{eq:OU_driven_SHO} in the main text, the time evolution of the statistical moments of the O-U-driven simple harmonic oscillator is given by
\begin{align}\label{eq:matrix_moments_appendix}
\frac{d}{dt}\begin{pmatrix}
\langle x^2 \rangle \\ \langle \dot{x}^2 \rangle\\ \langle \xi^2 \rangle \\\langle x\dot{x} \rangle \\ \langle x\xi \rangle \\ \langle \dot{x}\xi \rangle
\end{pmatrix}
=
\begin{pmatrix}
0 & 0 & 0 & 2 & 0 & 0 \\
0 & 0 & 0 & -2\omega_0^2 & 0 & 2 \\
0 & 0 & -2\omega_c & 0 & 0 & 0 \\
-\omega_0^2 & 1 & 0 & 0 & 1 & 0 \\
0 & 0 & 0 & 0 & -\omega_c & 1 \\
0 & 0 & 1 & 0 & -\omega_0^2 & -\omega_c
\end{pmatrix}
\begin{pmatrix}
\langle x^2 \rangle \\ \langle \dot{x}^2 \rangle \\ \langle \xi^2 \rangle \\ \langle x\dot{x} \rangle \\ \langle x\xi \rangle \\ \langle \dot{x}\xi \rangle
\end{pmatrix}
 + \begin{pmatrix}
0 \\ 0 \\ \sigma^2 \\ 0 \\ 0 \\ 0
\end{pmatrix}.
\end{align}
The mean energy is given by
\begin{equation}
    \langle E(t) \rangle = \frac{\omega_0^2 \langle x^2(t) \rangle + \langle \dot{x}^2(t) \rangle}{2},
\end{equation}
and its first derivative is
\begin{align}
    \langle \dot{E}(t) \rangle &= \frac{ \omega_0^2 \frac{d}{dt}\langle x^2(t) \rangle + \frac{d}{dt}\langle \dot{x}^2(t)\rangle}{2} \\
    &= -\omega_0^2\langle x\dot{x}\rangle + \langle \dot{x}\xi\rangle + \omega_0^2\langle x\dot{x}\rangle \\
    &= \langle \dot{x}\xi \rangle.
\end{align}
Since the noise is in statistical equilibrium, $\langle \xi^2\rangle = \sigma^2 / (2\omega_c)$ is constant, so determining $\langle \dot E(t) \rangle$ reduces to a two-dimensional problem.  The fifth row of~Eq.~\eqref{eq:matrix_moments_appendix} gives $\langle \dot x\xi\rangle = \frac{d}{dt}\langle x\xi\rangle + \omega_c\langle x\xi\rangle$.  We can therefore recast the time evolution of $\langle x\xi \rangle$ as a damped forced harmonic oscillator
\begin{equation}
    \frac{d^2}{dt^2}\left\langle x\xi \right\rangle + 2\omega_c \frac{d}{dt}\left\langle x\xi \right\rangle + (\omega_c^2 + \omega_0^2)\langle x\xi\rangle = \langle \xi^2\rangle.
\end{equation}
We then integrate to find $\langle E(t)\rangle$
\begin{align}
    \langle E(t)\rangle &= \int_{0}^t \bigg[\frac{d}{dt'}\langle x\xi(t') \rangle + \omega_c \langle x\xi(t') \rangle\bigg] dt'\\
    &= \frac{\sigma^2}{2(\omega_0^2 + \omega_c^2)}t + \frac{\sigma^2 (\omega_0^2 - \omega_c^2)}{2\omega_c(\omega_0^2 + \omega_c^2)^2} +\frac{\sigma^2e^{-\omega_c t}}{2\omega_c(\omega_0^2+\omega_c^2)^2}\big[ (\omega_c^2-\omega_0^2)\cos (\omega_0t) - 2\omega_0\omega_c \sin (\omega_0t)\big].
\end{align}

\section{Numerical simulation details}\label{appendix:numerical_simulation_detail}
The first-passage results shown in Figs.~\ref{fig:fptd_2_epsilons}~and~\ref{fig:ratio} are obtained by direct Monte Carlo integration of the finite-depth oscillator driven by the Ornstein--Uhlenbeck process.  For these simulations, we set the linear trap frequency to $\omega_0 = \pi$ and fix the resonant noise power $S_\xi(\omega_0) = S_0 = 2$.  Thus, we calculate the normalized escape time as $\tau_\text{esc} = T_1\dot E_\text{cold} / \Delta = T_1 /\Delta$.  Each trajectory starts from rest at the trap center, $x(0) = v(0) = 0$, and the initial noise value is drawn from its stationary distribution $N(0, \sigma^2/2\omega_c)$.  

We generated trajectories by two different methods in order to validate the solution method.  The first was the ``stability-optimized stochastic Runge--Kutta for It\^o'' (SOSRI) method as it has a strong convergence order 1.5~\cite{Rossler_2010}.  We used the standard $\mathtt{Julia\,  DifferentialEquations.jl}$ package~\cite{Julia_rackauckas2017differentialequations} to implement the first-passage simulations.

The second, implemented from first principles in $\mathtt{Python\;3}$, used a Strang splitting.  We used a symplectic velocity-Verlet update for the Hamiltonian part, with the O-U force entering both half-kicks and advanced between them by its exact one-step solution, $\xi \to e^{-\omega_c\Delta t}\xi + [\sigma^2(1 - e^{-2\omega_c\Delta t})/2\omega_c]^{1/2} N(0,1)$.   The advantage of symplectic schemes is their inherent noise-free stability at long times, which becomes important at the smallest $h$ values.  For example, $h = 6.45\times 10^{-5}$ requires $\sim 10^3$ oscillation periods before escape.  Over the course of these oscillations, the secular energy drift of a non-symplectic integrator would bias $T_1$ directly.  We used a fixed step $\Delta t = 0.001$ for the simulations whose results are shown in this manuscript.  Integrating identical noise realizations with $\Delta t=5\times10^{-4}$ instead of $10^{-3}$ changed $\tau_{\rm esc}$ by less than 1\%.  Replacing the energetic escape criterion $E\ge\Delta$ by the first crossing of $|x|=r$ or $|x|=1.5r$ changes $\tau_{\rm esc}$ by less than 1\% at $h=6.45\times10^{-5}$, and by up to 3\% and 13\% at $h=6.45\times10^{-4}$ and $6.45\times10^{-3}$.   Trajectories are integrated until escape is recorded, so the reported means are not biased by censoring.  For Figs.~\ref{fig:fptd_2_epsilons}~and~\ref{fig:ratio}, we generated $1.5\times 10^6$ independent trajectories at every $(h, \omega_c)$ pair, except for the leftmost points in Fig.~\ref{fig:ratio}(b) where $h=6.45\times 10^{-6}$, where we generated $10^5$.  For the more general colored noise points shown in Fig.~\ref{fig:colored_alpha} we also generated 16000 trajectories per point.  Error bars denote $1\sigma$ sampling errors, but are almost always too small to be seen in the plots.

Both methods yielded near-identical results, but the results shown in the text use the symplectic method as it is faster and therefore enables larger sample sizes.  Two independent analytic benchmarks of the simulations were verified.  First, in the harmonic limit $r\to\infty$, the simulated mean energy $\langle E(t) \rangle$ was compared with the exact result of~Eq.~\eqref{eq:sho_exact}, including the constant offset and the decaying oscillatory transient.  This confirms that the integrator reproduces the heating rate $\dot E_\text{cold} = S_0 / 2$.  Second, replacing the O-U noise by white noise of the same resonant power yields the exact martingale result.  

Analogous methods are used to simulate the system with the more general colored noise spectrum~Eq.~\eqref{eq:power_law_definition_with_cutoff}.  For further details on the simulations with colored noise, consult Appendix~\ref{appendix:colored_noise_generation}.

%%%%%%%%%%%%%%%%%%%%%%%%%%%%%%%%%%%%%%%%%%%%%%%%%%%%%%%%%%%%%%%%%%%%
%%%
%%%  Colored Noise
%%%
%%%%%%%%%%%%%%%%%%%%%%%%%%%%%%%%%%%%%%%%%%%%%%%%%%%%%%%%%%%%%%%%%%%%

\section{Colored noise generation}\label{appendix:colored_noise_generation}
\begin{figure}
    \centering
    \includegraphics[width=1\linewidth]{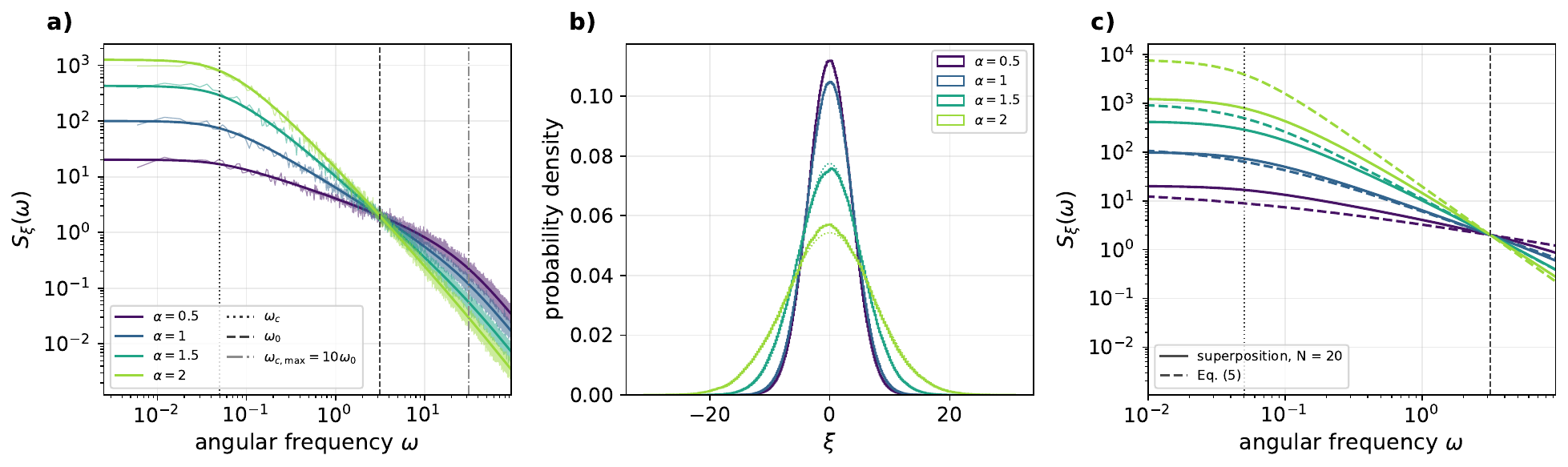}
    \caption{(a) Numerically evaluated power spectra of power-law noise generated by the O-U superposition used for Fig.~\ref{fig:colored_alpha} ($N=20$, $\omega_c=0.05$, $\omega_{c,\max}=10\omega_0$), compared with the exact spectrum of the superposition.  A single time series of duration $t=3\times10^4$ with $\Delta t=0.001$ is analyzed with Welch's method for each $\alpha$.  (b) Noise probability distribution compared with the expected Gaussian.  (c) Exact spectrum of the superposition (solid) compared with the soft-rolloff spectrum~Eq.~\eqref{eq:power_law_definition_with_cutoff} (dashed). The approximation is most faithful around $\alpha\sim1$.}
    \label{fig:power_spectra}
\end{figure}

There are several methods to generate colored noise whose spectrum is of the power-law form~Eq.~\eqref{eq:power_law_definition_with_cutoff}, such as the digital filtering method proposed in Ref.~\cite{PLASZCZYNSKI2007}.  For our simulation, we adapt the ideas originally put forward by McWhorter~\cite{Mcworther} to explain $1/f$ noise in semiconductors.  We simulate a power-law spectrum of the form~Eq.~\eqref{eq:power_law_definition_with_cutoff} (for $0 \leq \alpha \leq 2$) using a weighted sum of O-U processes
\begin{equation}
    \xi(t) = \sum_{i=0}^{N-1} \xi_i(t)
\end{equation}
where $\xi_i$ is an O-U process with parameters $\omega_{c_i}, \sigma_i$.  Each O-U process satisfies $d\xi_i = -\omega_{c_i}\xi_i dt + \sigma_i dW_i$, and has the usual Lorentzian power spectrum.  The $\omega_{c_i}$ are log-uniformly distributed according to:
\begin{equation}
    \omega_{c_i} = \exp \big(\log\omega_{c_\text{min}} + \lambda(\log\omega_{c_\text{max}} - \log\omega_{c_\text{min}})\big)
\end{equation}
where $\lambda = (i) / (N-1), \omega_{c_\text{max}} = 10\, \omega_0 , \omega_{c_\text{min} }=\omega_c$ of the synthesized process.  Beyond $\omega_{c_\text{max}}$ the spectrum decays as $1/\omega^2$.  In order to facilitate comparisons with a single O-U process with parameter $\omega_c$ and diffusion strength $\sigma = \sqrt{2(\omega_c^2 + \omega_0^2)}$, we choose the $\sigma_i$ such that the synthesized spectra have equal $S_0$ at $\omega_0$.  Thus
\begin{equation}
    \sigma_i = \mathtt{N}\sqrt{\omega_{c_i}^{(2-\alpha)}}
\end{equation}
where $\mathtt{N}$ is a normalization constant.  In these numerical experiments, whose results are shown in Fig.~\ref{fig:colored_alpha}, we used $N=20$, and $\omega_0 = \pi$.  We also used a time step of $0.001$, and simulated 16000 trajectories per point generated with the superposition method.    We show the numerically evaluated power spectra generated by this method in Fig.~\ref{fig:power_spectra}.  Panel $(c)$ illustrates the main text claim that this method is most faithful for $\alpha\sim 1$, and this is the primary origin behind the deviation between points generated by the superposition method with $\alpha = 2$ and pure O-U processes (Sec.~\ref{section:colored_noise_results}).  

We use this method as it is very light on memory because only the current state-vector must be stored (in contrast with some FIR filtering techniques) which allows us to handle very long trajectories ($> 10^7$ points).

\section{Amplitude-phase averaging}\label{appendix:amplitude_phase_averaging}
The transformation from $(x,\dot x)$ to amplitude-phase variables $(A,\phi)$,
\begin{equation}\label{eq:amp_phase_ansatz}
x = A\cos\Phi, \qquad \dot x = -A\,\omega_{\rm eff}(A)\sin\Phi,
\qquad \Phi = \omega_{\rm eff}(A)t+\phi(t),
\end{equation}
treats $A$ as slowly varying and $\Phi$ as rapidly varying, and introduces the amplitude-dependent frequency $\omega_{\rm eff}^2(u) = \omega_0^2(1-u)$ where we define $u\equiv 3A^2/4r^2$~\cite{Kougioumtzoglou_Spanos_2014, Kougioumtzoglou_2016}.  With negligible damping and weak noise, the response is approximately constant over one oscillation, and the averaged amplitude equation is~Eq.~\eqref{eq:averaged_SDE}.

For convenience, we can also write~Eq.~\eqref{eq:averaged_SDE} as $dA = B(A,t) dt + \sqrt{C(A,t)}dW$. This transformation into an It\^o equation is particularly useful as it transforms the additive colored noise in~Eq.~\eqref{eq:duffing} into a multiplicative white noise.  

This equation is fundamentally a weak-noise approximation.  For example, it does not capture the short-time oscillations (by assumption) in the mean energy that are observed when using the full equation and shown in Fig.~\ref{fig:energy}.  We note two qualitative features of~Eq.~\eqref{eq:averaged_SDE}.  Firstly, the drift term diverges at the left boundary $A=0$.  This boundary is categorized using Feller's criteria~\cite{feller_boundary} which show that it is an entrance boundary, which is a boundary that cannot be reached from the interior of the interval, but which is a valid initial starting point.  We note that, near the left boundary, our SDE reduces to a two-dimensional Bessel process~\cite{Martin_2011}.  Secondly, $\omega_\text{eff}$ vanishes at an amplitude $A_\text{max} = 2r/\sqrt{3}$.  Beyond this amplitude, the stiffness of the trap is negative, and it leads to unbounded response.  Ref.~\cite{Kougioumtzoglou_Spanos_2014} introduces a form of the conditional PDF of the response amplitude $P(A', t+\Delta t | A, t)$ that contains two conceptually different terms.  One represents the usual response PDF for values of $A<A_\text{max}$.  The other represents unbounded response with probability $1-\int_0^{A_\text{max}} dA'P(A', t+\Delta t | A, t)$.  Since we are interested in the first-passage time to that point, we do not need to consider dynamics beyond $A_\text{max}$.  

\subsection{Maximum energy of the averaged ansatz}

We compute the mechanical energy averaged over one oscillation of the amplitude-phase ansatz underlying the averaged equation~\eqref{eq:averaged_SDE}.  We show that it never reaches the trap depth $\Delta$ for any amplitude $0 \le A \le A_{\max}$.  Over one cycle the phase $\Phi$ is treated as uniformly distributed.  We use the following averages
\begin{equation}\label{eq:D_averages}
\langle \cos^2\Phi\rangle = \langle \sin^2\Phi\rangle = \tfrac{1}{2},
\qquad
\langle \cos^4\Phi\rangle = \tfrac{3}{8}.
\end{equation}
Throughout we use the rescaled squared amplitude $u \equiv 3 A^2/4r^2$, in terms of which
\begin{equation}
\frac{\omega_0^2 A^2}{2} = \frac{2\omega_0^2r^2}{3}\,u
= \frac{8\Delta}{3}\,u,
\label{eq:D_u_def}
\end{equation}
where $\Delta = \omega_0^2r^2/4$.  The cycle-averaged kinetic energy of~Eq.~\eqref{eq:amp_phase_ansatz} is
\begin{equation}
\langle T\rangle
= \tfrac{1}{2}\,A^2\omega_{\rm eff}^2(A)\,\langle\sin^2\Phi\rangle
= \frac{\omega_0^2 A^2}{4}\,(1-u).
\label{eq:D_kinetic}
\end{equation}
For the potential we write~Eq.~\eqref{eq:potential} in the form $V(x) = \tfrac{1}{2}\omega_0^2\left(x^2 - \tfrac{1}{2} x^4/r^2\right)$,
whose derivative reproduces the force in~Eq.~\eqref{eq:duffing}. Its cycle average over~Eq.~\eqref{eq:amp_phase_ansatz} is
\begin{align}
\langle V\rangle
&= \frac{\omega_0^2}{2}\left(A^2\langle\cos^2\Phi\rangle
 - \frac{ A^4}{2r^2}\,\langle\cos^4\Phi\rangle\right) \nonumber\\
&= \frac{\omega_0^2}{2}\left(\frac{A^2}{2} - \frac{3 A^4}{16r^2}\right)
= \frac{\omega_0^2 A^2}{4}\left(1 - \frac{u}{2}\right).
\label{eq:D_potential}
\end{align}
Adding~Eqs.~\eqref{eq:D_kinetic}~and~\eqref{eq:D_potential}, and using~Eq.~\eqref{eq:D_u_def} gives
\begin{equation}
\langle E\rangle(u)
= \frac{\omega_0^2 A^2}{2}\left(1 - \frac{3u}{4}\right)
= \frac{8\Delta}{3}\,u\left(1 - \frac{3u}{4}\right).
\label{eq:D_energy}
\end{equation}

Equation~\eqref{eq:D_energy} is a downward parabola in $u$ with $d\langle E\rangle/du \propto 1 - 3u/2$, maximized at $u^* = 2/3$, i.e., $A^* = \sqrt{8r^2/9}$, where
\begin{equation}
\langle E\rangle_{\max}
= \frac{8\Delta}{3}\cdot\frac{2}{3}\cdot\frac{1}{2}
= \frac{8\Delta}{9} < \Delta.
\label{eq:D_max}
\end{equation}
In particular, at the turning-point boundary $A = r$ ($u = 3/4$) the ansatz carries $\langle E\rangle = 7\Delta/8$, and at the dynamical endpoint $A = A_{\max}$ ($u = 1$), where $\omega_{\rm eff}$ vanishes and the motion is purely potential, it carries $\langle E\rangle = 2\Delta/3$. The mean energy of the ansatz is therefore non-monotonic in amplitude beyond $u = 2/3$ and bounded by $8\Delta/9$ on the entire interval $0 \le u \le 1$: no amplitude accessible to the averaged equation corresponds to the escape energy $E = \Delta$.

We make two observations.  First,~Eq.~\eqref{eq:D_energy} uses the first-order amplitude-dependent frequency $\omega_{\rm eff}^2 = \omega_0^2(1-u)$ at $u = O(1)$.  This is the internal convention of the closure, and~Eq.~\eqref{eq:D_max} is accordingly a statement about what the sinusoidal ansatz can represent, not about the true dynamics.  The exact conservative orbit at $E = \Delta$ (the separatrix) is a valid trajectory of~Eq.~\eqref{eq:duffing}, but it is strongly non-sinusoidal and lies outside the family of trajectories which satisfy~Eq.~\eqref{eq:amp_phase_ansatz}.  Second, the closure requirement creates an energy ceiling that sits below the barrier, which is one concrete measure of the $O(1)$ ambiguity, discussed in the main text.  This shows the difficulty in translating the energy criterion into an absorbing boundary on $A$.

\subsection{Escape-time distributions}

We set the absorbing boundary at $A_{\max}$ because it is the intrinsic endpoint of the averaged equation.  At $A=A_\text{max}$ the occupation probability satisfies $p(A_\text{max}, t)=0$.  Escape-time distributions are found using the backward Fokker--Planck equation, from which arises a hierarchical set of equations for the moments $T_n$ of the first-passage-time distribution, which is given by the generalized Pontryagin equation~\cite{Gardiner2009, VANVINCKENROYE2018}
\begin{equation}\label{eq:backwards_FPK}
    \frac{1}{2}C(x)\partial_x^2T_n(x) + B(x)\partial_xT_n(x) = -nT_{n-1}(x)
\end{equation}
with $T_0 = 1$.  This equation is solved directly by integration.  There are two constants of integration.  We impose $T_n(A_{\max})=0$ at the absorbing boundary and require the solution to remain finite at the entrance boundary $A=0$. 

For power-law spectra with a soft cutoff of the type given by~Eq.~\eqref{eq:power_law_definition_with_cutoff}, where $S_\xi(\omega) = S_0(\omega_0^\alpha + \omega_c^\alpha) / (\omega_c^\alpha + \omega^\alpha)$ so that $S_0 \equiv S_\xi(\omega_0)$, the Pontryagin equation~Eq.~\eqref{eq:backwards_FPK} yields 
\begin{equation}\label{eq:mfpt_integral_amplitude}
    T_1 = \frac{4\omega_0^2r^2}{3}\int_0^1\frac{(1-w)\ln (1/w)}{S_\xi(\omega_0\sqrt{1-w})}dw.
\end{equation}
We then obtain~Eq.~\eqref{eq:ratio_digamma}.  This shows that the cutoff location modifies the dynamics once $\omega_c^\alpha$ is not negligible against $\omega_0^\alpha$.  In the ordered weak-noise limit, and taking $\omega_c\to 0$, the mean first-passage time is given by
\begin{equation}\label{eq:amplitude_averaged_T1}
    \lim_{\omega_c\to 0}[T_1] = \frac{4\omega_0^2r^2}{3 S_0(2 + \alpha/2)}\bigg[\psi(3+\alpha/2) + \gamma_E\bigg].
\end{equation}
This behavior is consistent with observations in the previous sections, where it is seen that in the weak-noise limit the noise cutoff does not dramatically affect the escape dynamics.

Evaluating~Eq.~\eqref{eq:ratio_digamma} for the white-noise case ($\alpha=0$) gives $T_1 = \omega_0^2r^2/S_0 = 4\Delta/S_0$ which is exactly twice the exact martingale result~Eq.~\eqref{eq:martingale_anchor}.  We now argue that this discrepancy is an expected property of the choice of transformation variables, and that it largely cancels in the normalized ratio used below.    

This apparent error is not due to overall factors of the noise normalization or spectral convention.  Indeed, the averaged equation~Eq.~\eqref{eq:averaged_SDE} recovers the martingale result $T_1 = 2\Delta/S_0$ for the harmonic escape problem ($r\to\infty$ to a fixed energy threshold).

The factor of two arises because leading-order averaging is controlled only for $u \equiv 3A^2/4r^2 \ll 1$, while the first-passage time integrates through $u = O(1)$, and the identification of the energy threshold $E = \Delta$ with an amplitude is itself ambiguous.  Consistent with this origin, the overestimate depends on the choice of absorbing boundary.  Evaluating the solution of the Pontryagin equation at $\alpha=0$ with the boundary at $A=r$ rather than $A_\text{max}$ gives an overestimate of $13/8$ rather than $2$.  

\begin{figure}
    \centering
    \includegraphics[width=0.5\linewidth]{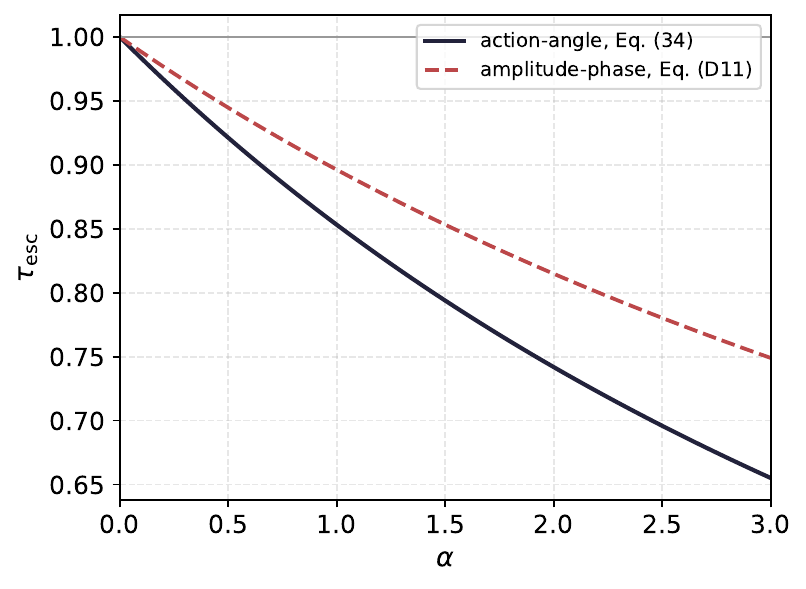}
    \caption{Comparison of the normalized escape time $\tau_\text{esc}(\alpha)$ in the weak-noise limit and taking $\omega_c\to 0 $, as calculated using action-angle averaging~Eq.~\eqref{eq:tau_pure_power}, and amplitude-phase averaging~Eq.~\eqref{eq:ratio_digamma} anchored by the martingale white noise result (i.e., divided by two), Eq.~\eqref{eq:amplitude_phase_heuristic}.}
    \label{fig:approximation_comparison}
\end{figure}

Because this closure error accumulates smoothly in $u$, it largely (though not exactly) cancels in the normalized ratio.  We therefore take from the averaged theory the \emph{shape} of the $\alpha$ dependence, fixing the $\alpha = 0$ value exactly by
~Eq.~\eqref{eq:martingale_anchor}.  This yields the normalized escape time
\begin{equation}\label{eq:amplitude_phase_heuristic}
    \tau_\text{esc} =  \frac{\omega_c^\alpha + \omega_0^\alpha\cdot\big(\frac{4\psi(3+\alpha / 2) + 4\gamma_E}{6 + 3\alpha/2}\big)}{\omega_c^\alpha + \omega_0^\alpha}
\end{equation}
which is shown in Fig.~\ref{fig:approximation_comparison} for $\omega_c = 0$.  We have shown that the amplitude-averaging method is a useful heuristic, but as it requires the white noise anchoring it is likely of limited theoretical use.  This method of averaging is compared to the action-angle averaging in Fig.~\ref{fig:approximation_comparison}, where it can be seen that as $\alpha$ increases, the former more severely overestimates the escape time.

\section{Harmonic truncation of $D(I)$}\label{appendix:fundamental_harmonic}

The action diffusion coefficient $D(I)$~Eq.~\eqref{eq:action_diffusion} samples the spectrum at all harmonics of the fundamental frequency of every orbit.  In the main text, we used Parseval's identity (which is exact) and the spectral-flatness-over-harmonics approximation to simplify $D(I)$~Eq.~\eqref{eq:action_diffusion_simplified}.  In this appendix, we evaluate this approximation and compare it to the full expression.  

The periodic orbits of the noise-free flow are known to be
\begin{equation}
    x_k = \frac{kr\sqrt{2}}{\sqrt{1+k^2}}\text{sn}\bigg(\frac{\omega_0 t}{\sqrt{1+k^2}}, k\bigg),
\end{equation}
where sn is a Jacobi elliptic function, and $k$ is the modulus satisfying $E/\Delta = 4k^2/(1+k^2)^2$.  The period associated with this motion is $\tilde T_k = 4\sqrt{1+k^2}\mathcal{K}(k) / \omega_0$, where $\mathcal{K}(k)$ is the complete elliptic integral of the first kind.  Making the identification $\theta = \omega t$, where $\omega = \pi\omega_0 / (2\mathcal{K}(k)\sqrt{1+k^2})$, we obtain
\begin{equation}
    x(k,\theta) = \frac{kr\sqrt{2}}{\sqrt{1+k^2}}\text{sn}\bigg(\frac{2\theta\mathcal{K}(k)}{\pi}, k\bigg).
\end{equation}
The function sn has a standard Fourier expansion, and thus we obtain
\begin{equation}
    x(k,\theta) = \frac{2\pi r\sqrt{2}}{\sqrt{1+k^2}\mathcal{K}}\sum_{n=0}^\infty \frac{q^{n+1/2}}{1-q^{2n+1}}\sin[(2n+1)\theta],
\end{equation}
where $q = \exp(-\pi\mathcal{K}'/\mathcal{K}), \; \mathcal{K}'(k)\equiv \mathcal{K}(\sqrt{1-k^2})$.  We can then immediately read off the Fourier coefficients of the position orbit
\begin{equation}
    x_n(k) = \frac{-i\pi r\sqrt{2}}{\sqrt{1+k^2}\mathcal{K}(k)} \frac{q(k)^{n/2}}{1-q(k)^{n}},
\end{equation}
for odd $n$.  The even $n$ components are identically 0.  From there, the last relation needed before $D(I)$ can be evaluated numerically is the relationship between the action $I$ and $k$.  We proceed as follows:
\begin{equation}
\begin{split}
    I &= \frac{1}{2\pi}\oint p\, dx \\
    &= \frac{2}{\pi}\int_0^{x_t} p \, dx \\
    &= \frac{2}{\pi}\int_0^{x_t} \frac{\omega_0}{\sqrt{2}r}\sqrt{(x_t^2-x^2)(x_+^2-x^2)}\, dx,
\end{split}    
\end{equation}
where we define, as is standard in the analysis of Duffing oscillators: $x_t = r\sqrt{1-\sqrt{1-E/\Delta}}$, and $x_+ = r\sqrt{1+\sqrt{1-E/\Delta}}$.  We also make the following identifications: $k = x_t / x_+, \: x_+ = \sqrt{2r^2 / (1+k^2)}, \: x = x_t\sin\phi$, and obtain:
\begin{equation}
\begin{split}
    I &= \frac{1}{2\pi}\oint p\, dx \\
    &= \frac{2}{\pi}\int_0^{\pi/2} \frac{2\omega_0k^2r^2}{(1+k^2)^{3/2}}\cos^2\phi\sqrt{1-k^2\sin^2\phi}\, d\phi \\
    &= \frac{4\omega_0r^2}{3\pi(1+k^2)^{3/2}}\bigg[(1+k^2)\mathcal{E}(k) - (1-k^2)\mathcal{K}(k)\bigg].
\end{split}    
\end{equation}
Here, we denote the complete elliptic integral of the second kind as $\mathcal{E}(k)$.

\begin{figure}
    \centering
    \includegraphics[width=0.5\linewidth]{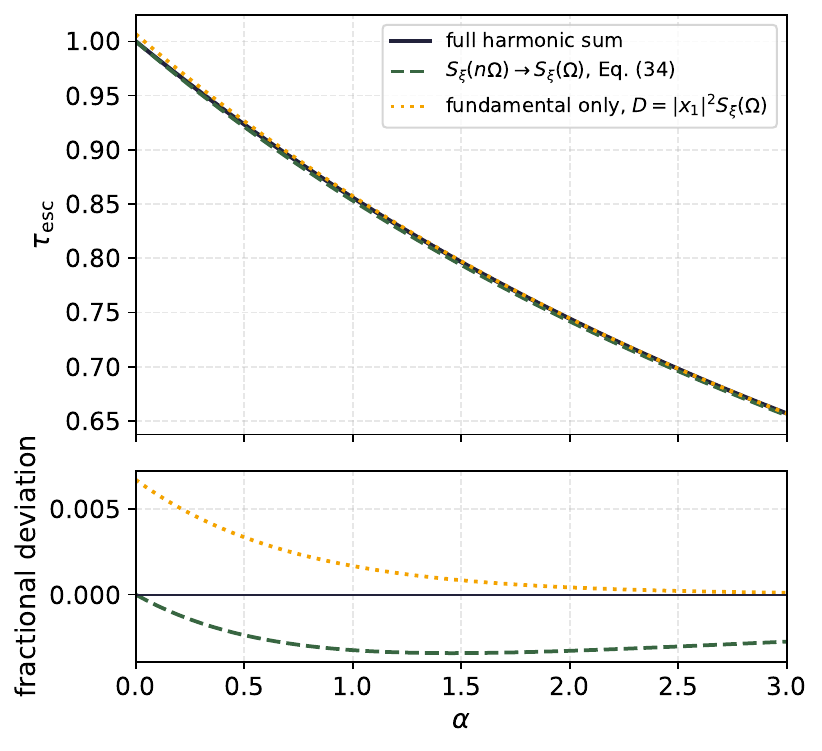}
    \caption{Normalized escape time $\tau_\text{esc}$ calculated from~Eq.~\eqref{eq:action_escape_time_complete} using different approximations of the action diffusion coefficient~Eq.~\eqref{eq:action_diffusion}, for a pure power-law spectrum with $\omega_c\to0$.  Upper panel: normalized escape time evaluated with 400 odd harmonics retained, $n=1,3,\ldots,799$ (solid), with the spectral-flatness replacement $S_\xi(n\Omega)\to S_\xi(\Omega)$ that yields~Eq.~\eqref{eq:tau_pure_power} (dashed), and with the literal fundamental-harmonic approximation $D=|x_1|^{2}S_\xi(\Omega)$ (dotted).  Lower panel: deviation from the full sum.  }
    \label{fig:fundamental_harmonic_approximation}
\end{figure}

We test two approximations in Fig.~\ref{fig:fundamental_harmonic_approximation}, and compare them with a numerical evaluation of the full $D(I)$ where we retain 400 odd harmonics, $n=1,3,\ldots,799$.  We show that retaining only the fundamental, where $D(I) = |x_1|^2S_\xi(\Omega)$, and making the spectral-flatness-over-harmonics approximation where $S_\xi(n\Omega) \to S_\xi(\Omega), \: \forall n$, both yield excellent approximations of $\tau_\text{esc}$.  This is because the fundamental carries almost all of the weight except within a small region around the separatrix, where the orbit becomes strongly non-sinusoidal.  

\section{Preparation of the initial state}\label{appendix:initial_state_prep}

\begin{figure}
    \centering
    \includegraphics[width=1\linewidth]{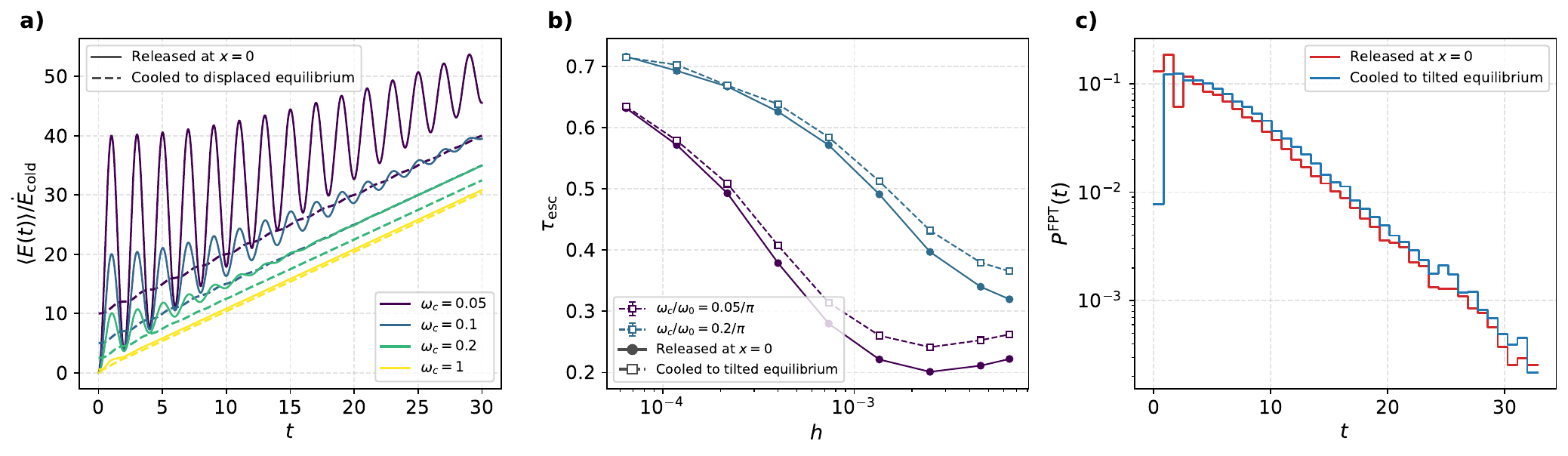}
    \caption{Preparation of the initial state.  Solid lines and filled symbols represent a particle released at rest at the untilted center with the force already present, as in the main text.  Dashed lines and open symbols represent a particle cooled to rest at the tilted equilibrium before release.  (a) Mean energy of the O-U-driven harmonic oscillator (counterpart of Fig.~\ref{fig:energy}), from Appendix~\ref{appendix:ou_sho}.  (b) Normalized escape time of the finite-depth oscillator for both preparations, driven by identical noise realizations.  (c) First-passage-time density at $h=6.45\times10^{-3}$ and $\omega_c/\omega_0=0.05/\pi$ (Fig.~\ref{fig:fptd_2_epsilons}(b)), released (red) and cooled (blue).  In panels (b) and (c), both preparations use only the realizations whose initial force leaves a minimum in the tilted well.}
    \label{fig:statePrep}
\end{figure}

Throughout the main text the particle starts at rest at $x=0$ while $\xi(0)$ is drawn from its stationary distribution.  For $\omega_c\ll\omega_0$, the force is nearly static during the initial oscillation and is already present when the particle is released at rest at the untilted center.  The alternative is a particle cooled to rest at the equilibrium of the instantaneously tilted potential before cooling is removed.

For the harmonic oscillator of Sec.~\ref{section:sho_ou} this equilibrium is $x(0)=\xi(0)/\omega_0^2$.  Repeating the calculation of Appendix~\ref{appendix:ou_sho} with these initial conditions gives the same heating rate and the offset
\begin{equation}
E_{\rm offset}^{\rm (c)} = E_{\rm offset} + \frac{\langle\xi^2\rangle}{2\omega_0^2} - \frac{\langle\xi^2\rangle}{(\omega_0^2+\omega_c^2)},\qquad \langle\xi^2\rangle=\frac{\sigma^2}{2\omega_c},
\end{equation}
which for $\omega_c\ll\omega_0$ is $E_{\rm offset}/2$, while the oscillatory transient is essentially absent (Fig.~\ref{fig:statePrep}(a)).  Half of the offset~Eq.~\eqref{eq:offset_energy} is the potential energy of the displaced equilibrium, common to both preparations.  The other half, together with the transient of Fig.~\ref{fig:energy}, is injected by the sudden release.

In the finite-depth trap the cooled particle starts at the stable minimum of $V(x)-\xi(0)x$.  When $|\xi(0)|r/\Delta>\chi_{\rm crit}$ the tilted well has no minimum and no particle can be prepared.  We compare both preparations on the remaining realizations, driven by identical noise.  Throughout the weak-noise regime, they give the same escape time, and separate only as $\chi$ grows, where the sudden release shortens the escape time (Fig.~\ref{fig:statePrep}(b)).  The released particle starts above the lowered saddle once $|\xi(0)|r/\Delta>(4/3)\sqrt{2/3}\approx1.09$, before the barrier disappears, and can then leave within half an oscillation.  The cooled particle has no such release energy (Fig.~\ref{fig:statePrep}(c)).  The short-time peak of Fig.~\ref{fig:fptd_2_epsilons}(b) is dominated by these large initial forces, including those for which no cooled particle can be prepared.  The leading weak-noise escape time is insensitive to the preparation.

 \end{document}